\documentstyle[amstex,amssymb,epsfig,12pt]{article}

\input{tcilatex}

\begin{document}

\bigskip 
\begin{titlepage}
\bigskip \begin{flushright}
WATPPHYS-TH01/09
\end{flushright}


\vspace{1cm}

\begin{center}
{\Large \bf {Abelian Higgs Hair for Rotating and Charged Black Holes    }}\\
\end{center}
\vspace{2cm}
\begin{center}
A. M. Ghezelbash$^{\dagger}${%
\footnote{%
EMail: amasoud@@avatar.uwaterloo.ca}} and
R. B. Mann$^\ddagger$
\footnote{
EMail: mann@@avatar.uwaterloo.ca} \\
$^{\dagger ,\ddagger}$Department of Physics, University of Waterloo, \\
Waterloo, Ontario N2L 3G1, CANADA\\
$^{\dagger}$Department of Physics, Alzahra University, \\
Tehran 19834, IRAN\\
\vspace{1cm}
\today\\
\end{center}

\begin{abstract}
We study the problem of vortex solutions in the background of rotating black holes
in both asymptotically flat and asymptoticlly anti de Sitter spacetimes. 
We demonstrate the Abelian Higgs field equations in the background 
of four dimensional Kerr, Kerr-AdS and Reissner-Nordstrom-AdS  black holes  
have vortex line solutions. These solutions, which have axial symmetry,  
are  generalization of the Nielsen-Olesen string. By numerically solving
the field equations in each case, we find that these black holes 
can support an Abelian Higgs field as hair. This situation holds even 
in the extremal case, and no flux-expulsion occurs.  We also compute the
effect of the self gravity of the  Abelian Higgs field show that the 
the vortex induces a deficit angle in the corresponding black hole metrics.

\end{abstract}
\end{titlepage}\onecolumn

\begin{center}
\bigskip 
\end{center}

\section{Introduction}

The conjecture that the only long-range information associated with the
endpoint of gravitational collapse is that of its total mass, angular
momentum and electric charge is referred to as the no-hair conjecture of
black holes \cite{Ruf}. Much work has been carried out over the years on
this conjecture, either upholding it in certain instances (e.g. scalar
fields \cite{Sud}) or challenging it in others, such painting Yang-Mills,
quantum hair \cite{Eli} or Nielsen-Olesen vortices \cite{Achu} on black
holes. In fact the uniqueness of the classical no-hair theorems is a
qualified uniqueness\cite{Cru}, incorporating additional criteria associated
with stability, non trivial topology, and the possibility of field
configurations on the horizon (referred to as `dressing'), and one must be
quite specific about what is meant by `hair' \cite{Achu}. The study of
Nielsen-Olesen vortices in the background of the charged black holes was
done in \cite{EM1},\cite{EM2},\cite{EM3} and \cite{EM4}.

Virtually all efforts in this area have been concerned with asymptotically
flat spacetimes, and it is only recently that extensions to other types of
asymptotia have been considered. Scalar fields minimally coupled to gravity
cannot provide hair for asymptotically de Sitter black holes \cite{Torii1},
but can do so if the spacetime is asymptotically anti de Sitter (AdS) \cite
{Torii2}. A solution to the $SU(2)$ Einstein-Yang-Mills equations that
describes a stable Yang-Mills hairy asymptotically AdS black hole has been
shown to exist \cite{Eli}. In a recent paper with Dehghani, we have shown
that the $U(1)$ Higgs field equations have a vortex solution in both four
dimensional AdS spacetime \cite{Deh} and AdS-Schwarzschild backgrounds \cite
{Deh2}. \ 

In this paper we extend our investigation of possible vortex hair for
non-asymptotically flat black holes to include rotation and charge.
Specifically, we seek numerical solutions of the Abelian-Higgs field
equations in the four dimensional rotating Kerr-AdS and
Reissner-Nordstrom-AdS black hole backgrounds. Although an analytic or
approximate solution to these equations appears to be intractable, we
confirm by numerical calculation that Kerr-AdS and Reissner-Nordstrom-AdS
black holes could support a long range vortex (or cosmic string) as a form
of stable hair. \ This is the first demonstration that rotating black holds
can carry Abelian Higgs hair, for both the asymptotically flat and
asymptotically AdS cases. \ We also show (to first order in the
gravitational coupling) that the effect of the vortex on a rotating black
hole is to create a deficit angle through the spacetime, analogous to its
non-rotating counterpart. We consider the question of flux-expulsion by
extremal black holes and find that in the case of extremal Kerr-AdS, Kerr
and Reissner-Nordstrom-AdS black holes, an extremal horizon is indeed
pierced by the vortex, and its flux is not expelled from the black holes.
This not only confirms earlier computations of non-expulsion of vortex lines 
\cite{EM3,EM4} for the extremal Reissner-Nordstrom case, but also proves the
non-expulsion phenomenon to the extremal rotating case in asymptotically AdS
and asymptotically flat spacetimes. Finally, we consider the dependence of
the characteristic vortex on the rotation parameter of the black hole and
find that it is qualitatively the same its dependence on the magnitude of
the cosmological constant: an increase in both causes a decrease in the
thickness of the core.

\bigskip

In section two, we solve numerically the Abelian-Higgs equations in the
Kerr-AdS background for different values of the cosmological constant and
black hole rotation parameter. In section three, we consider the
Abelian-Higgs equations in the limiting case of \ Kerr-AdS background with
cosmological constant to be zero, yielding the vortex equation in the Kerr
black hole background. In section four, we repeat this calculation for a
Reissner-Nordstrom-AdS background for different values of the black hole
charge. In the limiting case of Reissner-Nordstrom-AdS black hole with
cosmological constant to be zero, yielding Reissner-Nordstrom black hole,
our solution is in good agreement with the previous obtained results \cite
{EM4}. In section five, by an analytical discussion, we obtain the effect of
the vortex self gravity on the Kerr black hole to the first order of
gravitational constant. We establish that the effect of the vortex is to
induce a deficit angle in the Kerr background and find the deficit angle in
terms of vortex fields. Then in section six, by studying the behaviour of
the string energy-momentum tensor, we find the effect of the vortex self
gravity on the Kerr-AdS and Reissner-Nordstrom-AdS black hole backgrounds.
We give some closing remarks in the final section. \ 

\section{Abelian Higgs Vortex in Kerr$-$AdS Black Hole}

The Abelian Higgs Lagrangian is

\begin{equation}
{\cal L}(\Phi ,A_{\mu })=-\frac{1}{2}({\cal D}_{\mu }\Phi )^{\dagger }{\cal D%
}^{\mu }\Phi -\frac{1}{16\pi }{\cal F}_{\mu \nu }{\cal F}^{\mu \nu }-\xi
(\Phi ^{\dagger }\Phi -\eta ^{2})^{2}  \label{Lag}
\end{equation}
where $\Phi $ is a complex \ scalar Klein-Gordon field, ${\cal F}_{\mu \nu }$
is the field strength of \ the electromagnetic field $A_{\mu }$ and ${\cal D}%
_{\mu }=\nabla _{\mu }+ieA_{\mu }$ in which $\nabla _{\mu }$ is the
covariant derivative in a spacetime with metric $g_{\mu \nu }$. We employ
Planck units $G=\hbar =c=1$ which implies that the Planck mass is equal to
unity, and write the Kerr-AdS black hole metric in the analogue of
Boyer-Lindquist coordinates 
\begin{equation}
ds^{2}=-\frac{\Delta _{r}}{\rho ^{2}}(dt-\frac{a}{\Sigma }\sin ^{2}\theta
d\varphi )^{2}+\frac{\rho ^{2}}{\Delta _{r}}dr^{2}+\frac{\rho ^{2}}{\Delta
_{\theta }}d\theta ^{2}+\frac{\Delta _{\theta }\sin ^{2}\theta }{\rho ^{2}}%
(adt-\frac{r^{2}+a^{2}}{\Sigma }d\varphi )^{2}  \label{kerradsmetr1}
\end{equation}
\qquad where 
\begin{equation}
\Delta _{r}=(r^{2}+a^{2})(1+\frac{r^{2}}{l^{2}})-2mr  \label{deltar}
\end{equation}

\begin{equation}
\begin{array}{c}
\Delta _{\theta }=1-\frac{a^{2}\cos ^{2}\theta }{l^{2}} \\ 
\Sigma =1-\frac{a^{2}}{l^{2}}
\end{array}
\label{para}
\end{equation}
and $\rho ^{2}=r^{2}+a^{2}\cos ^{2}\theta .$ The parameter $m$ is related to
the mass of the black hole $M$ by 
\[
M=\frac{ml}{(1-\frac{a^{2}}{l^{2}})^{2}} 
\]
and $a$ to the angular momentum $J$ by 
\[
J=\frac{am}{(1-\frac{a^{2}}{l^{2}})^{2}} 
\]
The cosmological constant $\Lambda $ $\ $is equal to $\frac{-3}{l^{2}}.$ The
metric (\ref{kerradsmetr1}) is valid only where $\left| a\right| <l$ and is
singular where $\left| a\right| =l.$ If \ $m>0,$ $\Delta _{r\text{ }}$has at
most two real roots and the event horizon is located at $r=r_{H},$ the
largest real root of the equation $\Delta _{r}=0.$ For some values of the
parameters $a,m,l,$\ the two real roots coincide in which case the Kerr-AdS
black hole is extremal.{\large \ }If roots are not real, then the spacetime
described by the metric (\ref{kerradsmetr1}) is a naked singularity. As in
the non-rotating case \cite{Deh2}, we define the real fields $X(x^{\mu
}),\omega (x^{\mu }),P_{\mu }(x^{\nu })$ via the following equations

\begin{equation}
\begin{tabular}{l}
$\Phi (x^{\mu })=\eta X(x^{\mu })e^{i\omega (x^{\mu })}$ \\ 
$A_{\mu }(x^{\nu })=\frac{1}{e}(P_{\mu }(x^{\nu })-\nabla _{\mu }\omega
(x^{\mu }))$%
\end{tabular}
\label{XPomegadef}
\end{equation}
from which we can rewrite the Lagrangian ( \ref{Lag}) and the equations of
motion in terms of these fields as 
\begin{equation}
{\cal L(}X,P_{\mu })=-\frac{\eta ^{2}}{2}(\nabla _{\mu }X\,\nabla ^{\mu
}X+X^{2}P_{\mu }P^{\mu })-\frac{1}{16\pi e^{2}}F_{\mu \nu }F^{\mu \nu }-\xi
\eta ^{4}(X^{2}-1)^{2}  \label{Lag2}
\end{equation}

\begin{equation}
\begin{tabular}{l}
$\nabla _{\mu }\nabla ^{\mu }X-XP_{\mu }P^{\mu }-4\xi \eta ^{2}X(X^{2}-1)=0$
\\ 
$\nabla _{\mu }F^{\mu \nu }-4\pi e^{2}\eta ^{2}P^{\nu }X^{2}=0$%
\end{tabular}
\label{eqmo2}
\end{equation}
where $F^{\mu \nu }=\nabla ^{\mu }P^{\nu }-\nabla ^{\nu }P^{\mu }$ is the
field strength of the corresponding gauge field $P^{\mu }$.

Provided the field $\omega $ is not single valued, the resultant solutions
contain physical information and are referred to as vortex solutions \cite
{NO}. \ The requirement that $\Phi $ be single-valued implies that the line
integral of $\omega $ over any closed loop is $\pm 2\pi n$ where $n$ is an
integer. Through such a closed loop the flux of electromagnetic field $\Phi
_{H\text{ \ }}$is quantized with quanta $2\pi /e.$

We seek a vortex solution for the Abelian Higgs Lagrangian (\ref{Lag2}) in
the background of Kerr-AdS black hole. This solution can be interpreted as a
string piercing the black hole (\ref{kerradsmetr1}). Considering the static
case of winding number $N$ with the gauge choice, 
\begin{equation}
P_{\mu }(r,\theta )=(0;0,0,NP(r,\theta ))  \label{Pgauge}
\end{equation}
and $X=X(r,\theta )$, we rescale 
\begin{equation}
\varkappa \rightarrow \frac{\varkappa }{\sqrt{\xi }\eta }  \label{rescale}
\end{equation}
where $\varkappa =r,l,m,$ thereby obtaining

\begin{equation}
\begin{array}{c}
\frac{(r^{2}+l^{2})(r^{2}+a^{2})-2mrl^{2}}{l^{2}(r^{2}+a^{2}\cos ^{2}\theta )%
}\frac{\partial ^{2}X}{\partial r^{2}}+2\frac{2r^{3}+r(l^{2}+a^{2})-ml^{2}}{%
l^{2}(r^{2}+a^{2}\cos ^{2}\theta )}\frac{\partial X}{\partial r}+\frac{%
l^{2}-a^{2}\cos ^{2}\theta }{l^{2}(r^{2}+a^{2}\cos ^{2}\theta )}\frac{%
\partial ^{2}X}{\partial \theta ^{2}}+\cot \theta \frac{l^{2}+2a^{2}-3a^{2}%
\cos ^{2}\theta }{l^{2}(r^{2}+a^{2}\cos ^{2}\theta )}\frac{\partial X}{%
\partial \theta } \\ 
+\frac{(a-l)^{2}(a+l)^{2}(r^{4}+(r^{2}+a^{2}\cos ^{2}\theta
)(a^{2}+l^{2})-a^{4}\cos ^{4}\theta +2mrl^{2})}{l^{2}\sin ^{2}\theta
(r^{2}+a^{2}\cos ^{2}\theta )(a^{2}\cos ^{2}\theta
-l^{2})(r^{4}+r^{2}(a^{2}+l^{2})+a^{2}l^{2}-2mrl^{2})}N^{2}XP^{2}-\frac{1}{2}%
X(X^{2}-1)=0
\end{array}
\label{eqx}
\end{equation}


\begin{equation}
\begin{array}{c}
-\frac{(r^{2}+l^{2})(r^{2}+a^{2})-2mrl^{2}}{l^{2}(r^{2}+a^{2}\cos ^{2}\theta
)}\frac{\partial ^{2}P}{\partial r^{2}}-2\frac{(a^{2}+r^{2})(r^{5}+2a^{2}%
\cos ^{2}\theta r^{3}+ml^{2}r^{2}+a^{4}\cos ^{4}\theta r-a^{2}ml^{2}\cos
^{2}\theta )}{l^{2}(r^{2}+a^{2}\cos ^{2}\theta )^{3}}\frac{\partial P}{%
\partial r}+\frac{a^{2}\cos ^{2}\theta -l^{2}}{l^{2}(r^{2}+a^{2}\cos
^{2}\theta )}\frac{\partial ^{2}P}{\partial \theta ^{2}} \\ 
+\cot \theta \frac{r^{4}(l^{2}-2a^{2}+a^{2}\cos ^{2}\theta
)+2a^{2}r^{2}(l^{2}\cos ^{2}\theta +a^{2}\cos ^{4}\theta -2a^{2}\cos
^{2}\theta )+4a^{2}ml^{2}\sin ^{2}\theta r+a^{4}\cos ^{2}\theta (l^{2}\cos
^{2}\theta +a^{2}\cos ^{4}\theta -2a^{2})}{l^{2}(r^{2}+a^{2}\cos ^{2}\theta
)^{3}}\frac{\partial P}{\partial \theta } \\ 
+\alpha PX^{2}=0
\end{array}
\label{eqp}
\end{equation}
for the equations of motion (\ref{eqmo2}), where $\alpha =\frac{4\pi e^{2}}{%
\xi }.$ The equations (\ref{eqx}) and (\ref{eqp}) in the special cases of \ $%
a=0$ and $a=m=0$ reduce to the equations of motion of the vortex in the
Schwarzschild-AdS and AdS backgrounds respectively \cite{Deh2},\cite{Deh}.

We emphasize that even in the simplest case of an Abelian Higgs vortex in
asymptotically flat spacetime ((\ref{eqx}) and (\ref{eqp}) in the limit $%
l\rightarrow \infty ,$\ $a=0$), no solution that is everywhere analytic has
been found.\ Indeed, even for $m=0$ \ no exact analytic solutions are known
for equations (\ref{eqx}) and (\ref{eqp}).{\large \ }Furthermore, no vortex
solutions have ever been obtained for rotating black hole spacetimes. We now
proceed with a numerical search for the existence of vortex solutions for
the above coupled non linear partial differential equations.

First, we consider the {\it thin }string with winding number one, in which
one can assume $m>>1.$ Thicker vortices and larger winding numbers will be
discussed later in this section. Employing the ansatz

\begin{equation}
P(r,\theta )=P(R),\text{ \ }X(r,\theta )=X(R)  \label{PXcylsym}
\end{equation}
where $R=r\sin \theta $,\ we obtain the following equations:

\begin{equation}
\begin{array}{c}
-\frac{%
a^{2}(R^{4}+r^{4}-3r^{2}R^{2}-R^{2}l^{2})-r^{4}(l^{2}+R^{2})+2rml^{2}R^{2}}{%
l^{2}(r^{4}+a^{2}(r^{2}-R^{2}))}\frac{d^{2}X}{dR^{2}}-\frac{%
a^{2}(r^{4}-7r^{2}R^{2}+4R^{4})-r^{4}(4R^{2}+l^{2})+2R^{2}ml^{2}r}{%
Rl^{2}(r^{4}+a^{2}(r^{2}-R^{2}))}\frac{dX}{dR} \\ 
-\frac{1}{2}(X^{3}-X)+f(r,\theta )N^{2}XP^{2}=0
\end{array}
\label{eqx2}
\end{equation}

\begin{equation}
\frac{r^{2}(a^{2}\cos ^{4}\theta -l^{2})-\sin ^{2}\theta
(r^{4}+a^{2}r^{2}-2mrl^{2}+a^{2}l^{2})}{l^{2}(r^{2}+a^{2}\cos ^{2}\theta )}%
\frac{d^{2}P}{dR^{2}}+\frac{g(r,\theta )}{l^{2}\sin \theta (r^{2}+a^{2}\cos
^{2}\theta )^{3}}\frac{dP}{dR}-\alpha PX^{2}=0  \label{eqp2}
\end{equation}
where the functions $f(r,\theta )$ and $g(r,\theta )$ are given by, 
\begin{equation}
f(r,\theta )=\frac{%
r^{2}(a-l)^{2}(a+l)^{2}(a^{4}R^{2}r^{2}-a^{4}R^{4}+a^{2}r^{6}+a^{2}r^{4}l^{2}-a^{2}R^{2}l^{2}r^{2}-2r^{5}ml^{2}+r^{8}+l^{2}r^{6})%
}{R^{2}l^{2}\{r^{4}+a^{2}(r^{2}-R^{2})\}\{r^{2}(a^{2}-l^{2})-a^{2}R^{2}\}%
\{(r^{2}+l^{2})(r^{2}+a^{2})-2mrl^{2}\}}  \label{f}
\end{equation}
\begin{equation}
\begin{array}{c}
g(r,\theta )=-2\sin ^{2}\theta r^{7}+(6\cos ^{4}\theta
a^{2}+l^{2}-5a^{2}\cos ^{2}\theta -2a^{2})r^{5}-2ml^{2}\sin ^{2}\theta r^{4}+
\\ 
\{-4a^{4}\cos ^{2}\theta (1+\cos ^{2}\theta )+2a^{2}l^{2}\cos ^{2}\theta
+6a^{4}\cos ^{6}\theta \}r^{3}-2a^{2}ml^{2}(1+3\cos ^{4}\theta -4\cos
^{2}\theta )r^{2} \\ 
+a^{4}\cos ^{4}\theta (l^{2}-2a^{2}+2a^{2}\cos ^{4}\theta -a^{2}\cos
^{2}\theta )+2a^{4}ml^{2}\sin ^{2}\theta \cos ^{2}\theta
\end{array}
\label{g}
\end{equation}

When $a=0,$ the vortex solutions of the Abelian Higgs equations in flat
spacetime (without a black hole) satisfy the $l\rightarrow \infty $ limit of
equations up to errors which are proportional to $\frac{mR^{2}}{r^{3}}%
\approx \frac{m}{r^{3}}$\cite{Deh}$.$ These errors are very tiny far from
the black hole horizon, whereas near the horizon $r\approx r_{H}=2m,$ they
are of the order of $\frac{1}{m^{2}},$ which is negligible for large mass
black holes. This suggests that a string vortex solution could be painted to
the horizon of a Schwarzschild black hole, and numerical calculations \cite
{Achu} have indeed shown the existence of vortex solutions of the Abelian
Higgs equations in this background.

The situation is somewhat different for finite $l$ . For\ $a=0$ (and $l\neq
0 $) we have shown in a previous paper \cite{Deh} that the Abelian Higgs
equations of motion \ in the background of Anti-de-Sitter spacetime ( (\ref
{eqx2}) and (\ref{eqp2}) in the limit of \ $a=0$ and $m=0$) have vortex
solutions (denoted by $X_{0}$ and $P_{0}$) with core radius $R\approx O(1)$
. The functions $X_{0}$ and $P_{0}$ satisfy eqs. (\ref{eqx2}) and (\ref{eqp2}%
) up to errors which are proportional to $\frac{mR^{2}}{r^{3}}\approx \frac{m%
}{r^{3}}.$ Although these errors go to zero far from the black hole, for a
large mass black hole, $\ r\approx r_{H}\approx m^{1/3}$ , the term $\frac{m%
}{r^{3}}$ is at least of \ the order of unity, and so the possibility of
obtaining a string vortex solution for finite $l$ in this background is
unclear. \ 

By numerically solving eqs. (\ref{eqx2},\ref{eqp2}) in the $a=0$ case we
have shown that vortex solutions exist on, near and far from the horizon of
the AdS-Schwarzschild black hole for various winding numbers and different
values of $l$ \cite{Deh2}. As in the asymptotically flat case the results
indicate that increasing the winding number yields a greater vortex
thickness. Furthermore as $l$ decreases the black hole becomes completely
covered by a vortex of decreasingly large winding number. Also, for a vortex
with definite winding number, the string core decreases with decreasing $l,$
but the ratio of string core to the size of the black hole horizon
increases. The $X$ \ and $P$ fields less rapidly approach their respective
maximum and minimum values at larger angles as $l$ decreases.

For finite values of the rotation parameter $a$, equations (\ref{eqx2}) and (%
\ref{eqp2}) which are equivalent to equations (\ref{eqx}) and (\ref{eqp})
are considerably more complicated than in the $a=0$ case. To obtain
numerical solutions of \ (\ref{eqx}) and (\ref{eqp}) outside the black hole
horizon we must first select appropriate boundary conditions. At large
distances from the horizon physical considerations motivate a clear choice:
we demand that the $a\neq 0$\ solutions approach the solutions of the vortex
equations in pure AdS spacetime given in ref. \cite{Deh}. This means that we
demand $X\rightarrow 1$ and $P\rightarrow 0$ as $R$ goes to infinity. On the
symmetry axis of the string and beyond the radius of horizon $r_{H}$, i.e. $%
\theta =0$ and $\theta =\pi $, we take $X\rightarrow 0$ and $P\rightarrow 1$
. Finally, on the horizon, we initially take $X=0$ and $P=1.$

We then employ a polar grid of points $(r_{i},\theta _{j}),$ where $r$ goes
from $r_{H}$ to some large value of $r$ ( $r_{\infty }$) which is much
greater than $r_{H}$ and $\theta $ runs from $0$ to $\pi .$ \ We use the
finite difference method and rewrite the non linear partial differential
equation (\ref{eqx}) and (\ref{eqp}) as 
\begin{equation}
A_{ij}X_{i+1,j}+B_{ij}X_{i-1,j}+C_{ij}X_{i,j+1}+D_{ij}X_{i,j-1}+E_{ij}X_{i,j}=F_{ij}
\label{findiffxx}
\end{equation}
\begin{equation}
A_{ij}^{\prime }P_{i+1,j}+B_{ij}^{\prime }P_{i-1,j}+C_{ij}^{\prime
}P_{i,j+1}+D_{ij}^{\prime }P_{i,j-1}+E_{ij}^{\prime }P_{i,j}=F_{ij}^{\prime }
\label{findiffxp}
\end{equation}
where $X_{ij}=X(r_{i},\theta _{j})$ and $P_{ij}=P(r_{i},\theta _{j}).$ For
the interior grid points and horizon grid points, the coefficients $%
A_{ij},...,F_{ij\text{ }}^{\prime }$ can be straightforwardly determined
from the corresponding continued differential equations (\ref{eqx}) and (\ref
{eqp}). The form of the coefficients is somewhat complicated, and we so
relegate them to an appendix.

Using the well known successive overrelaxation method \cite{num} for the
above mentioned finite difference equations, we obtain the values of $X$ and 
$P$ fields inside the grid, which we denote them by $X^{(1)}$ and $P^{(1)}$.
Then by calculating the $r$-gradients of $X$ and $P$ just outside the
horizon and iterating the finite difference equations on the horizon, we get
the new values of $X$ and $P$ fields on the horizon points. Then these new
values of $X$ and $P$ fields are used as the new boundary condition on the
horizon for the next step in obtaining the values of $X$ and $P$ fields
inside the grid which could be denoted by $X^{(2)}$ and $P^{(2)}$. Repeating
this procedure, the value of the each field in the $(n+1)$ -th iteration is
related to the $n$-th iteration by

\begin{equation}
X_{ij}^{(n+1)}=X_{ij}^{(n)}-\omega \frac{\zeta _{ij}^{(n)}}{E_{ij}^{(n)}}
\label{sor}
\end{equation}
\begin{equation}
P_{ij}^{(n+1)}=P_{ij}^{(n)}-\omega \frac{\varsigma _{ij}^{(n)}}{%
E_{ij}^{\prime (n)}}  \label{sorp}
\end{equation}
where the residual matrices $\zeta _{ij}^{(n)}$ and $\varsigma _{ij}^{(n)}$
are the differences between the left and right hand sides of the equations (%
\label{findiffx}\ref{findiffxx}) and (\ref{findiffxp}) respectively,
evaluated in the $n$-th iteration and $\omega $ is the overrelaxation
parameter. The iteration is performed many times to some value $n=K,$ such
that $\sum_{i,j}\left| X_{ij}^{(K)}-X_{ij}^{(K-1)}\right| <\varepsilon $ and 
$\sum_{i,j}\left| P_{ij}^{(K)}-P_{ij}^{(K-1)}\right| <\varepsilon $ for a
given error $\varepsilon $. It is a matter of trial and error to find the
value of $\omega $ that yields the most rapid convergence.

\begin{figure}[tbp]
\begin{center}
\epsfig{file=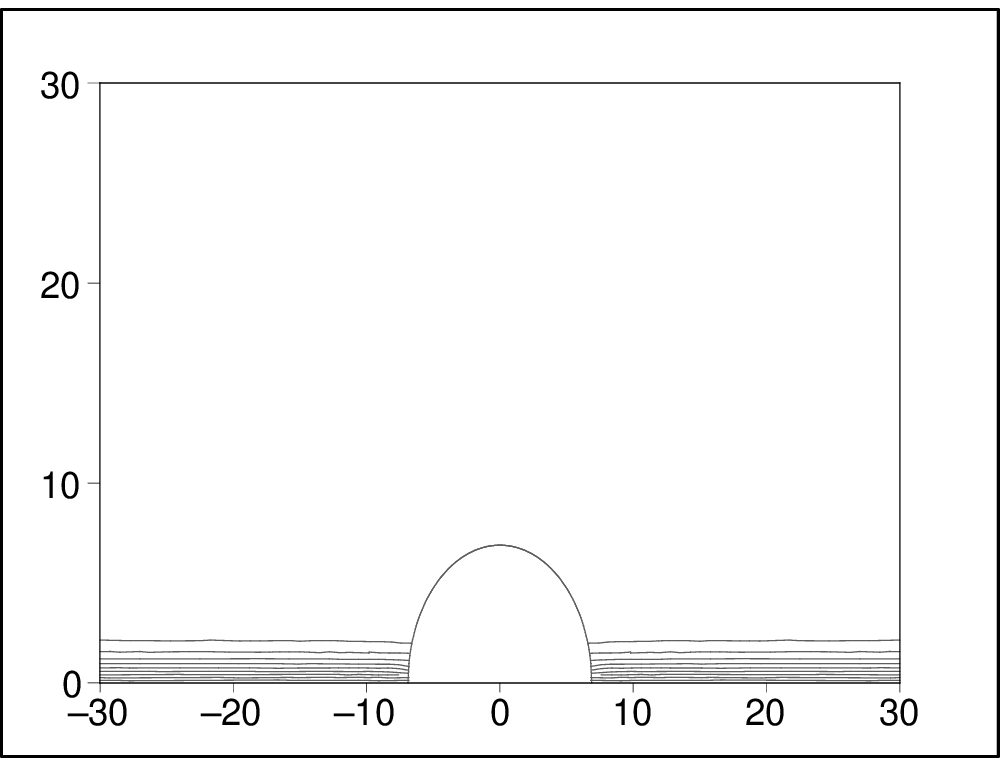,width=0.4\linewidth} \epsfig{file=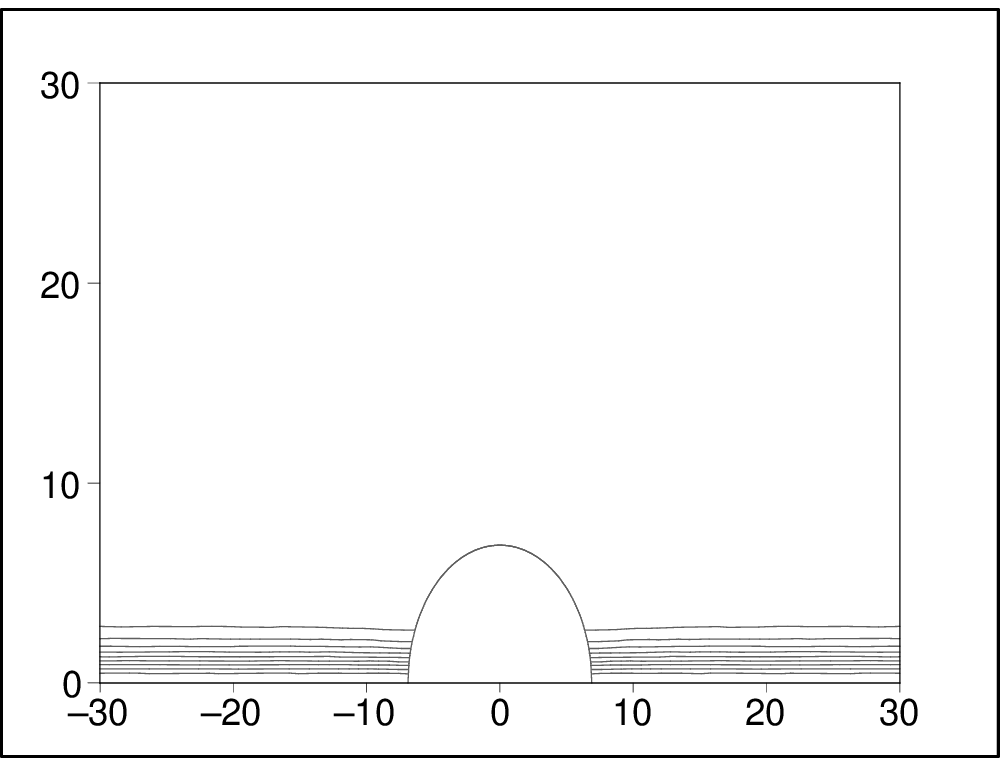,width=0.4%
\linewidth}
\end{center}
\caption{{}$X$ and $P$ contours ($X$ and $P$ increase from $0.1$ to $0.9$
upward and downward respectively) in the background of Kerr-AdS black hole
with $l=5$ and $a=0$. The mass of black hole is equal to $10$ and horizon is
located in $r_{H}=6.89.$}
\label{fig1}
\end{figure}

\begin{figure}[tbp]
\begin{center}
\epsfig{file=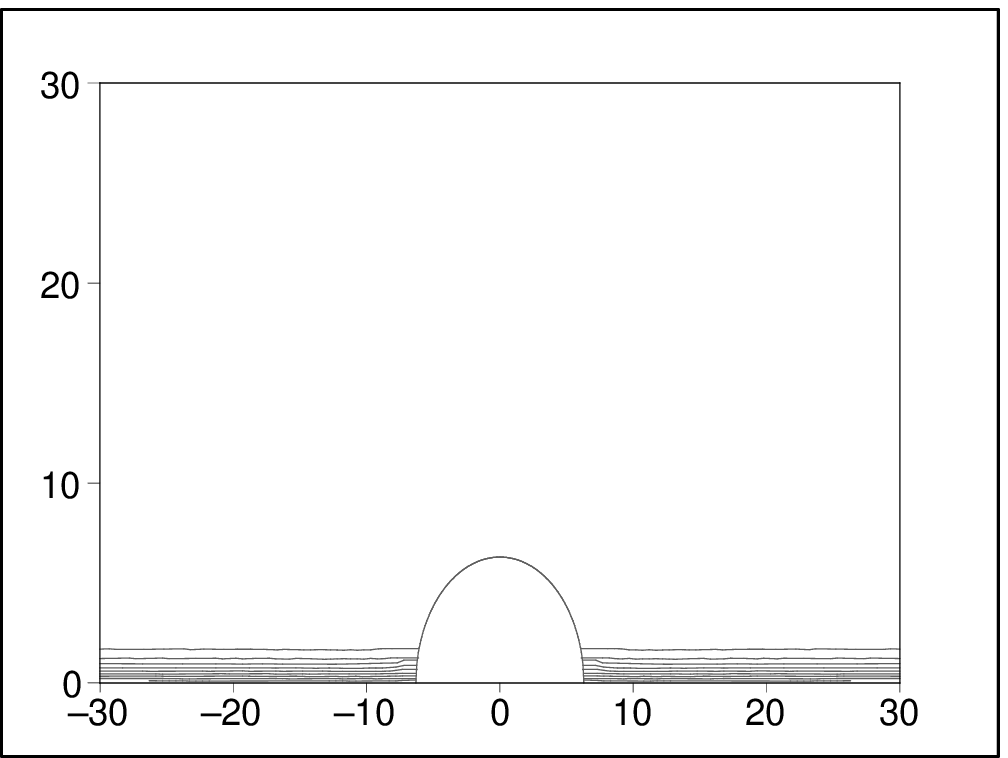,width=0.4\linewidth} \epsfig{file=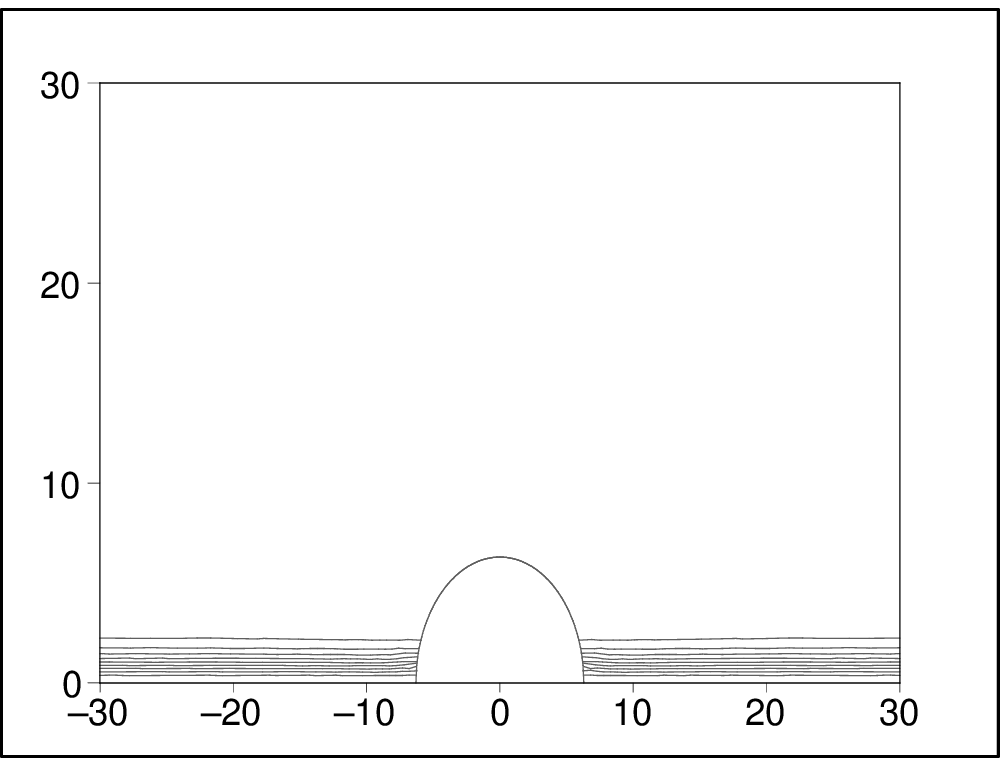,width=0.4%
\linewidth}
\end{center}
\caption{{}$X$ and $P$ contours ($X$ and $P$ increase from $0.1$ to $0.9$
upward and downward respectively) in the background of Kerr-AdS black hole
with $l=5$ and $a=3$. The mass of black hole is equal to $10$ and horizon is
located in $r_{H}=6.30.$}
\label{fig2}
\end{figure}

\begin{figure}[tbp]
\begin{center}
\epsfig{file=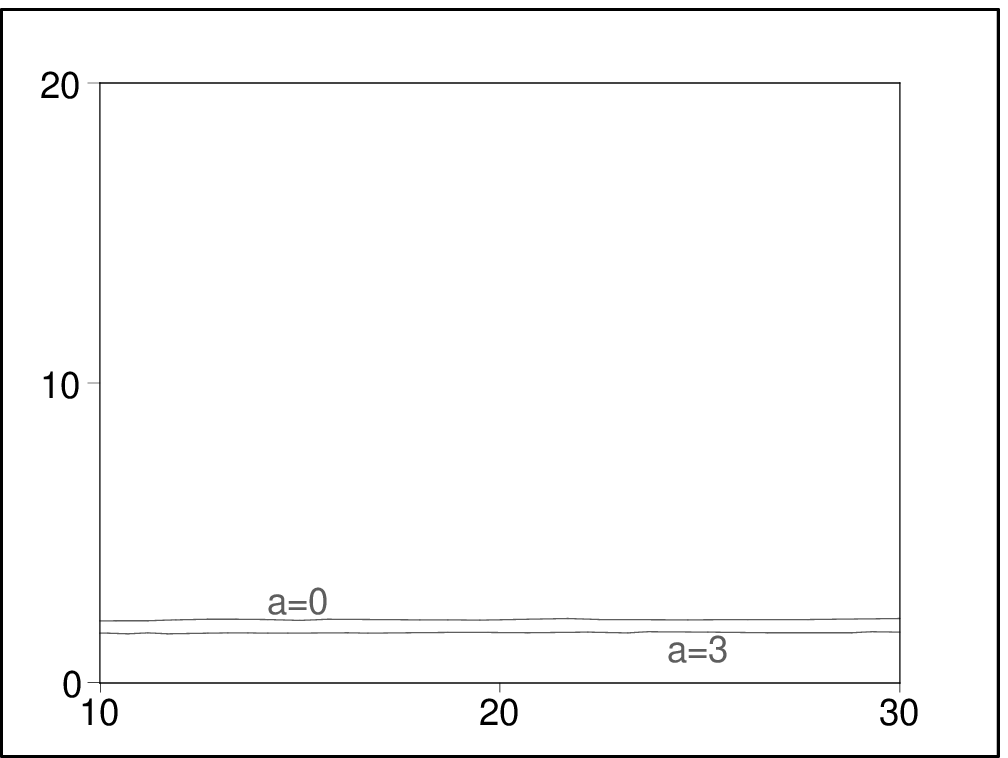,width=0.4\linewidth} %
\epsfig{file=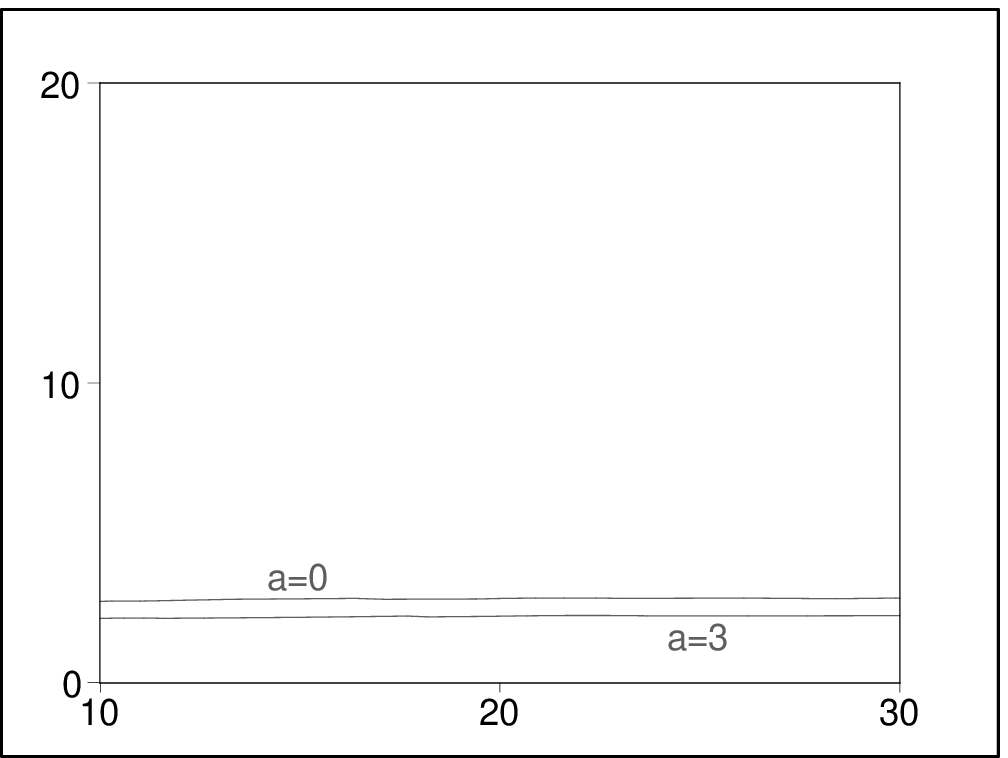,width=0.4\linewidth}
\end{center}
\caption{{}$X=0.9$ and $P=0.1$ contours in the presence of $l=5$ Kerr-AdS
black hole for different values of the rotation parameter $a=0,3$. The black
hole horizon is located off to the left of the figure at $r_{H}=6.89,6.30$
respectively.}
\label{fig3}
\end{figure}

\begin{figure}[tbp]
\begin{center}
\epsfig{file=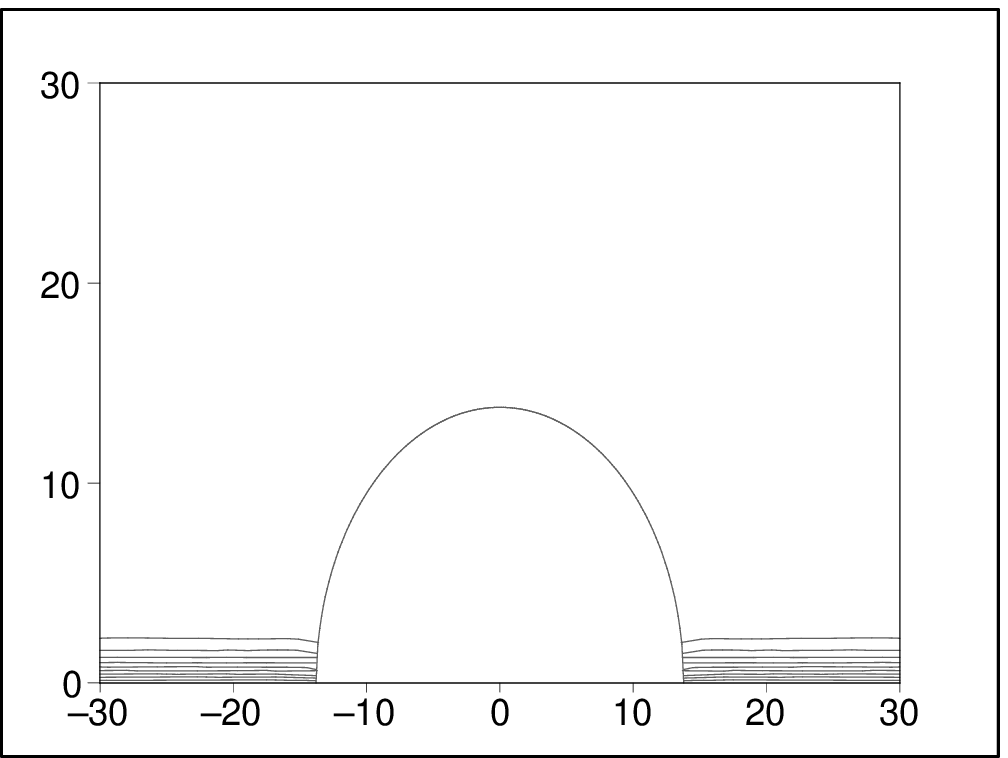,width=0.4\linewidth} %
\epsfig{file=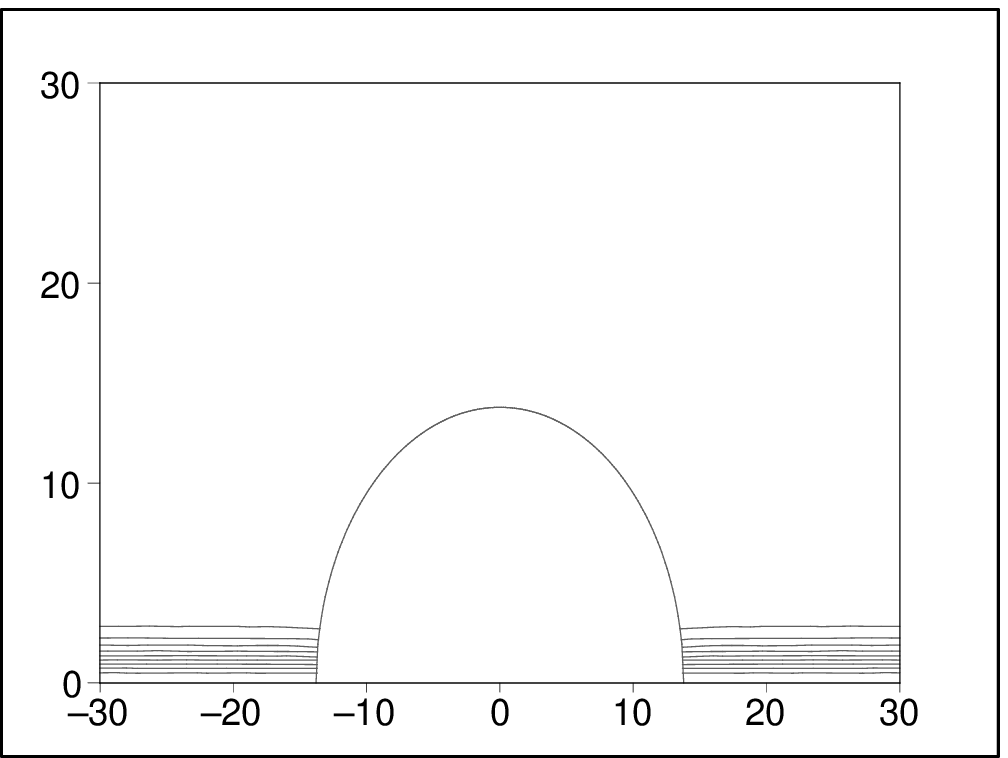,width=0.4\linewidth}
\end{center}
\caption{{}$X$ and $P$ contours ($X$ and $P$ increase from $0.1$ to $0.9$
upward and downward respectively) in the background of Kerr-AdS black hole
with $l=10$ and $a=0$. The mass of black hole is equal to $20$ and horizon
is located in $r_{H}=13.79.$}
\label{fig4}
\end{figure}

\begin{figure}[tbp]
\begin{center}
\epsfig{file=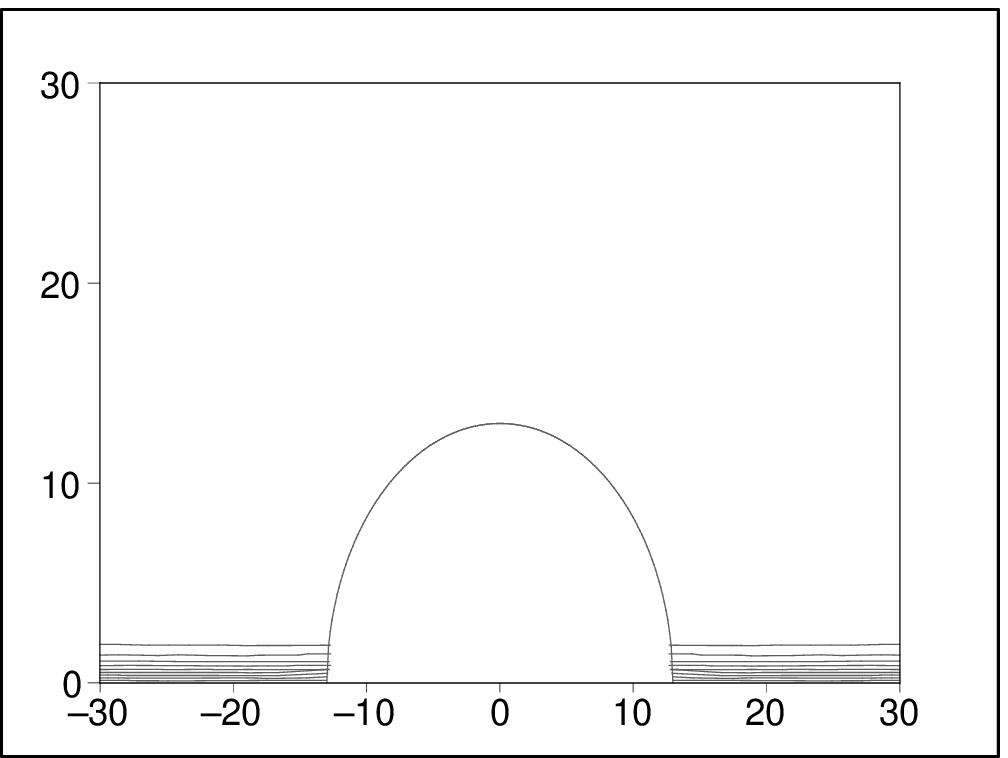,width=0.4\linewidth} %
\epsfig{file=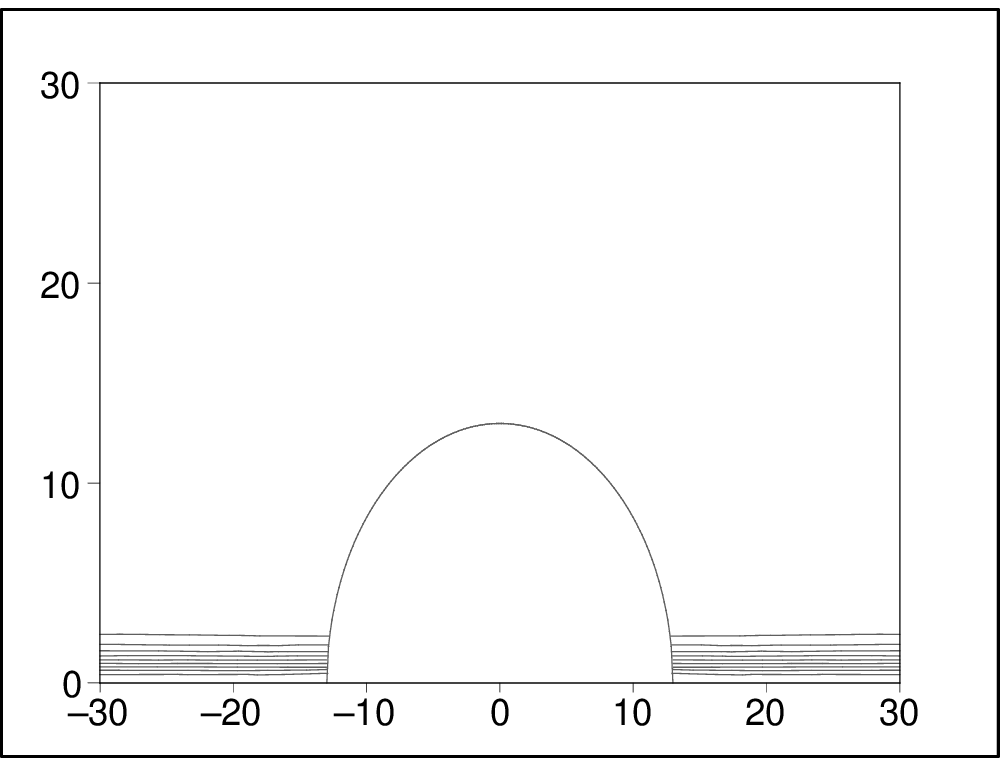,width=0.4\linewidth}
\end{center}
\caption{{}$X$ and $P$ contours ($X$ and $P$ increase from $0.1$ to $0.9$
upward and downward respectively) in the background of Kerr-AdS black hole
with $l=10$ and $a=5$. The mass of black hole is equal to $20$ and horizon
is located in $r_{H}=12.98.$}
\label{fig5}
\end{figure}

\begin{figure}[tbp]
\begin{center}
\epsfig{file=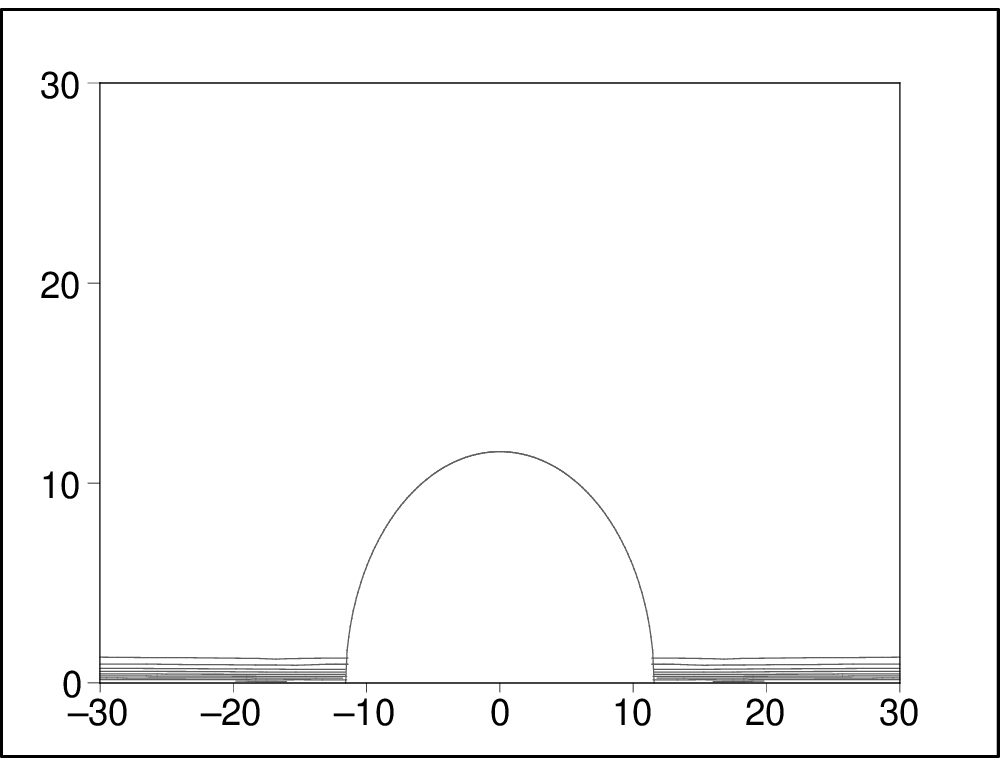,width=0.4\linewidth} %
\epsfig{file=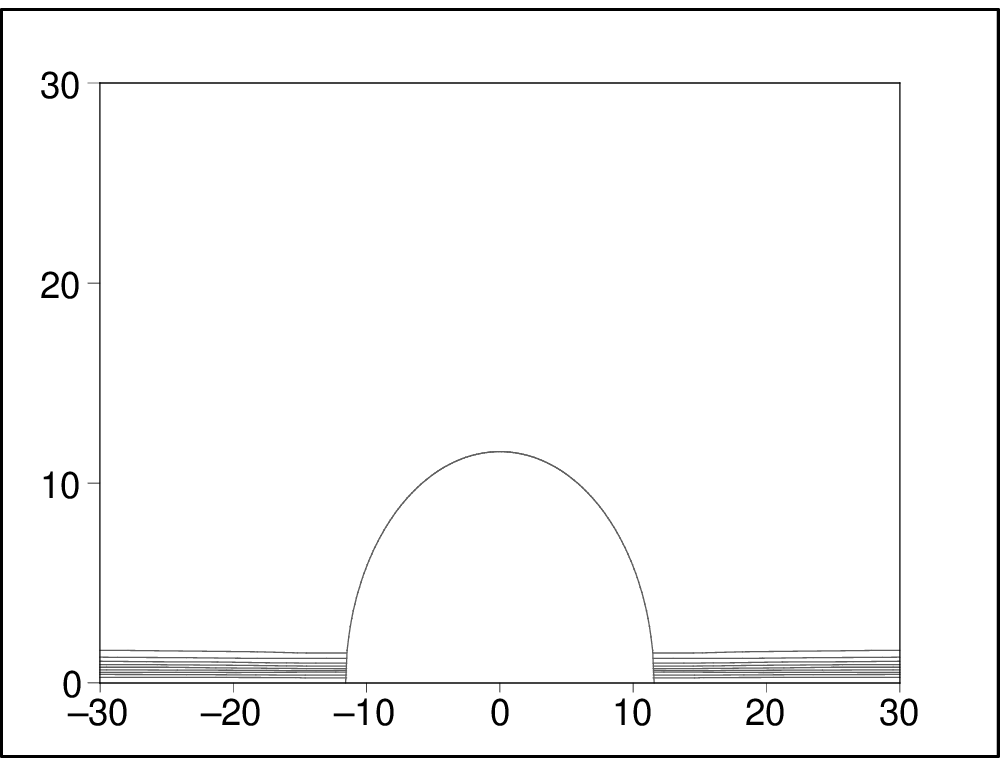,width=0.4\linewidth}
\end{center}
\caption{{}$X$ and $P$ contours ($X$ and $P$ increase from $0.1$ to $0.9$
upward and downward respectively) in the background of Kerr-AdS black hole
with $l=10$ and $a=8$. The mass of black hole is equal to $20$ and horizon
is located in $r_{H}=11.57.$}
\label{fig6}
\end{figure}

Some typical results of this calculation are displayed in figures (\ref{fig1}%
), (\ref{fig2}) and (\ref{fig4}),(\ref{fig5}), (\ref{fig6}) for different
values of $\ l=5$ and $l=10$, respectively.\ In the first case, we consider
both $a=0$ and $a=3$ and in the second case, we consider $a=0,5,8.$ \ For $%
l=5$, the black hole mass is taken to be the constant value $m=10$ whereas
for $l=10$ the mass is taken to be $m=20.$ \ In $l=5,$ the horizon is
located in $r_{H}=6.89$ for $a=0$ and $r_{H}=6.30$ for $a=3.$ Also, for the
case of $l=10,$ the horizon for different values of $a=0,5,8$ are $%
r_{H}=13.79,12.98,11.57$ respectively. For these two values of $l$, the
black holes are non-extremal for all values of $a$.\ The diagrams (\ref{fig1}%
) and (\ref{fig4}) when $a=0$ are exactly the same as the results of \cite
{Deh2} for the AdS-Schwarzschild black hole. We notice that by increasing
the rotation parameter $a$ from $0$ to $3,$ the string core decreases
slightly in the $l=5$ Kerr-AdS black hole.\ Figure (\ref{fig3}) shows
explicitly the string core decreasing as the rotation parameter increases.

The physical reason for the core size decreasing with increasing rotation
parameter $a$\ is due to a decrease in the horizon radius $r_{H}$. This
quantity is given by the root of equation (\ref{deltar}) and decreases
slowly as the parameter $a$\ increases from$\ 0$\ to its maximum value.\ On
the other hand, since we know from previous work that the string thickness
compared to horizon radius drastically changes by changing unbounded
physical quantities like the winding number or black hole mass, we expect
that string core changes a little with changing the bounded black hole
parameters like the rotation parameter $a$. Since the effect of increasing $%
l $\ (when $a=0$) is the same as the effect of decreasing $a$ on the value
of the horizon size\ $r_{H},$ we expect commensurate changes in the horizon
size.\ \ Indeed this is what we find: the core thickness increases with
decreasing parameter $a$, analogous to what happens when the value of $l$ is
increased for $a=0$.

\begin{figure}[tbp]
\begin{center}
\epsfig{file=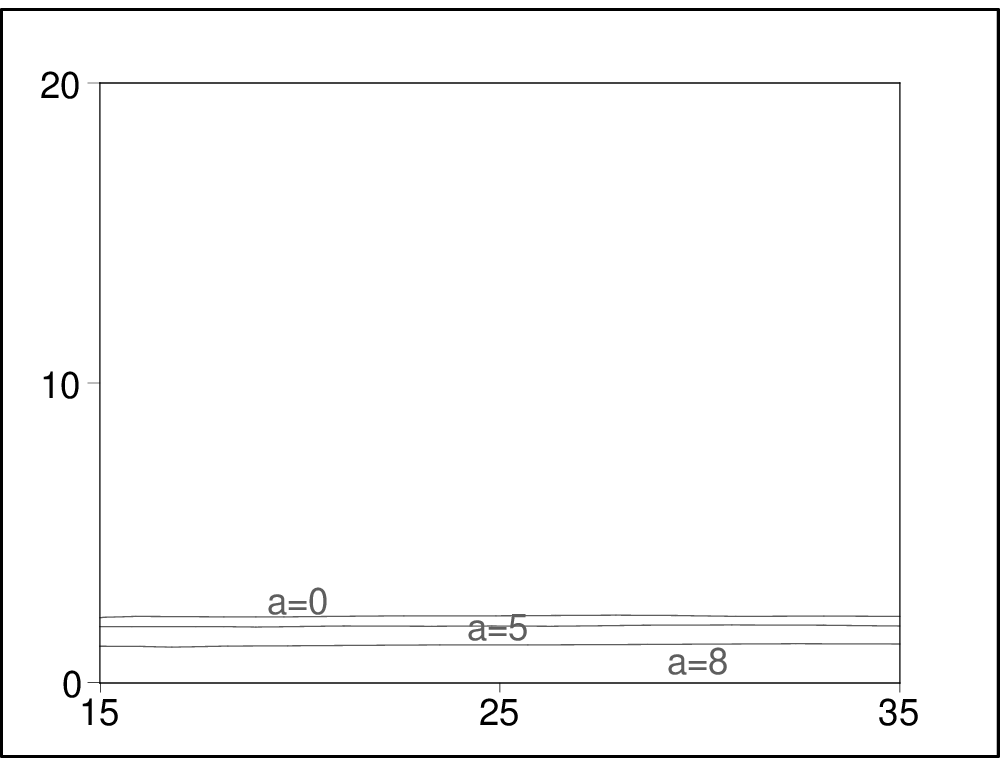,width=0.4\linewidth} %
\epsfig{file=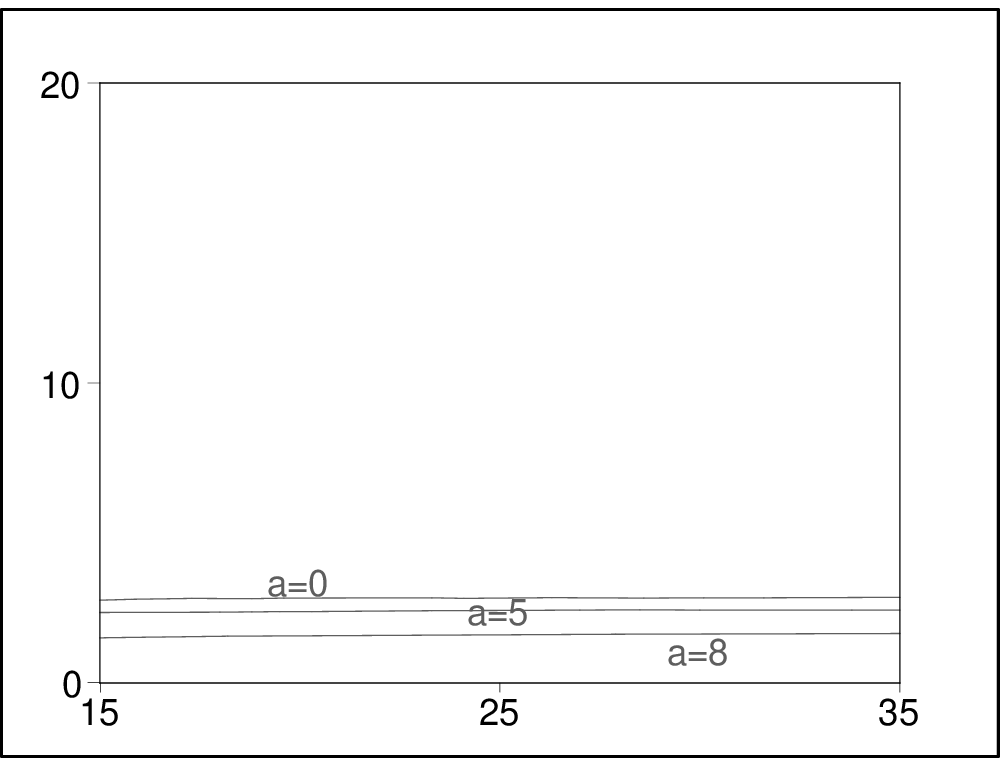,width=0.4\linewidth}
\end{center}
\caption{{}$X=0.9$ and $P=0.1$ contours in the presence of $l=10$ Kerr-AdS
black hole for different values of the rotation parameter $a=0,5,8$. The
black hole horizon is located off to the left of the figure at $%
r_{H}=13.79,12.98,11.57$ respectively.}
\label{fig7}
\end{figure}

Figure (\ref{fig7}) illustrates for $l=10$\ a similar decrease of the string
core as $a$\ increases from $0$\ to $5$\ and then to $8,$\ confirming again
our expectation. Our numerical results agree with the above statement
relating the horizon radius and different physical quantities of the black
hole. Similar calculations for the string with larger winding numbers in a
Kerr-AdS black hole with definite parameters $l,a$, show that increasing the
winding number yields a greater vortex thickness.

\bigskip

For the extremal Kerr-AdS black hole with cosmological parameters $l=10,$\
mass $m=11$\ and rotation parameter $a=7.939195$\ we have calculated the
vortex fields. In this case the horizon radius is located at $r_{H}=5.11015.$%
\ We find that the behaviour of the vortex $X$\ and $P$\ fields are the same
as the non-extremal cases. The contours of the $X$\ and $P$\ fields attach
to the horizon in the same way they attach to the horizon of non extremal
black holes presented in the figures \ref{fig1},\ref{fig2},\ref{fig4},\ref
{fig5},\ref{fig6}. Since the results are so similar to the non-extremal
cases we do not present them here. The crucial point is that in the extremal
case, no flux-expulsion is occurred in the thin vortex configuration.

\section{Abelian Higgs Vortex in a Kerr Black Hole}

\begin{figure}[tbp]
\begin{center}
\epsfig{file=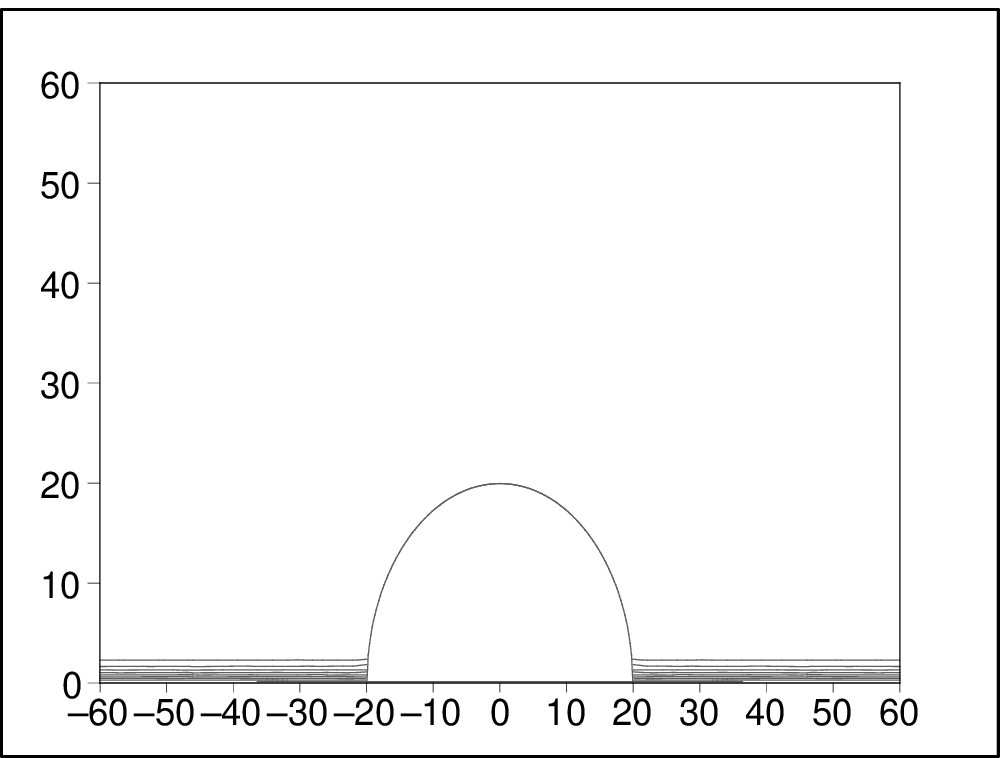,width=0.4\linewidth} \epsfig{file=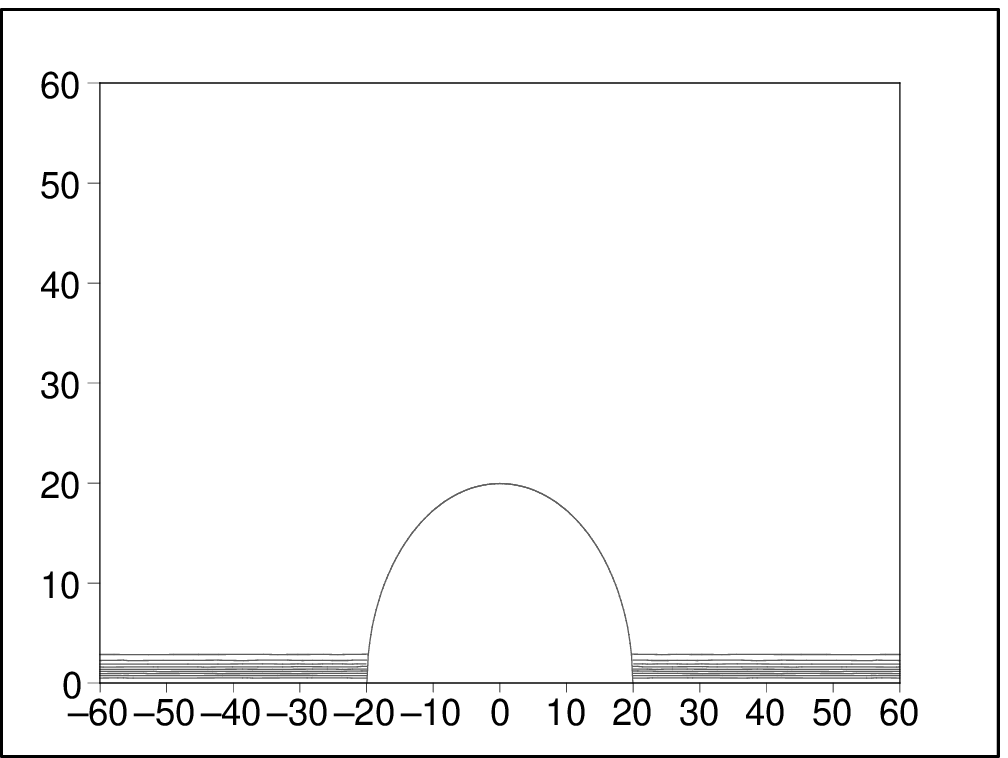,width=0.4%
\linewidth}
\end{center}
\caption{{}$X$ and $P$ contours ($X$ and $P$ increase from $0.1$ to $0.9$
upward and downward respectively) for a string in the presence of
Schwarzschild black hole as a limiting case of a Kerr black hole where $a=0$%
. The mass of the black hole is equal to $10$ and horizon is located in $%
r_{H}=20.$}
\label{fig8}
\end{figure}

\begin{figure}[tbp]
\begin{center}
\epsfig{file=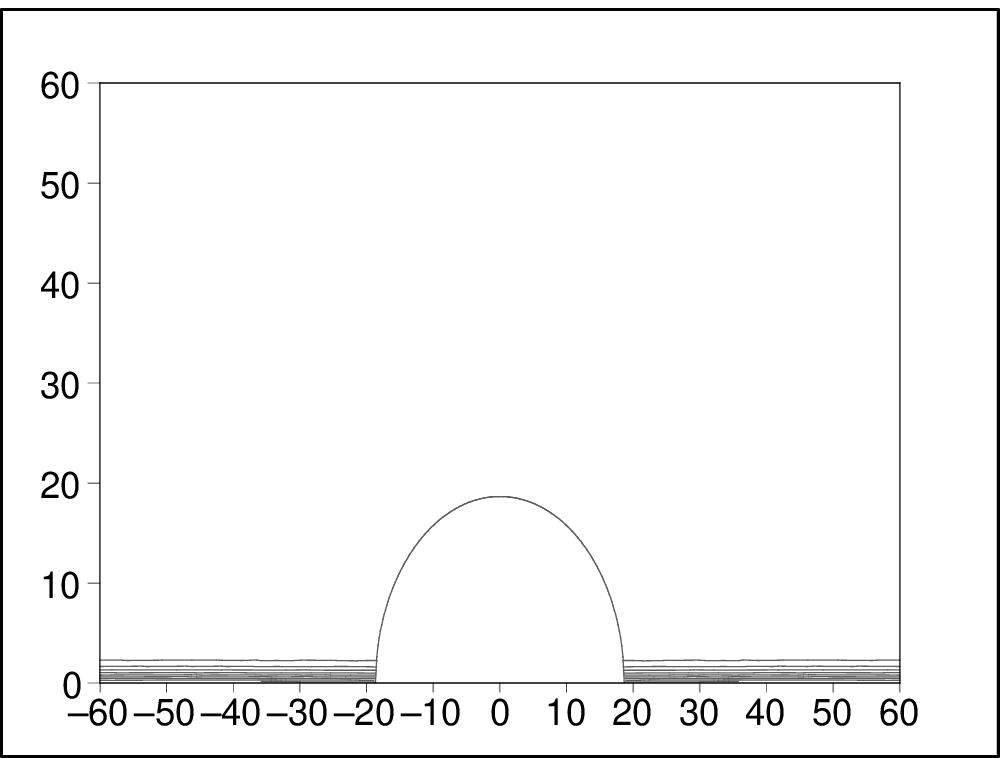,width=0.4\linewidth} \epsfig{file=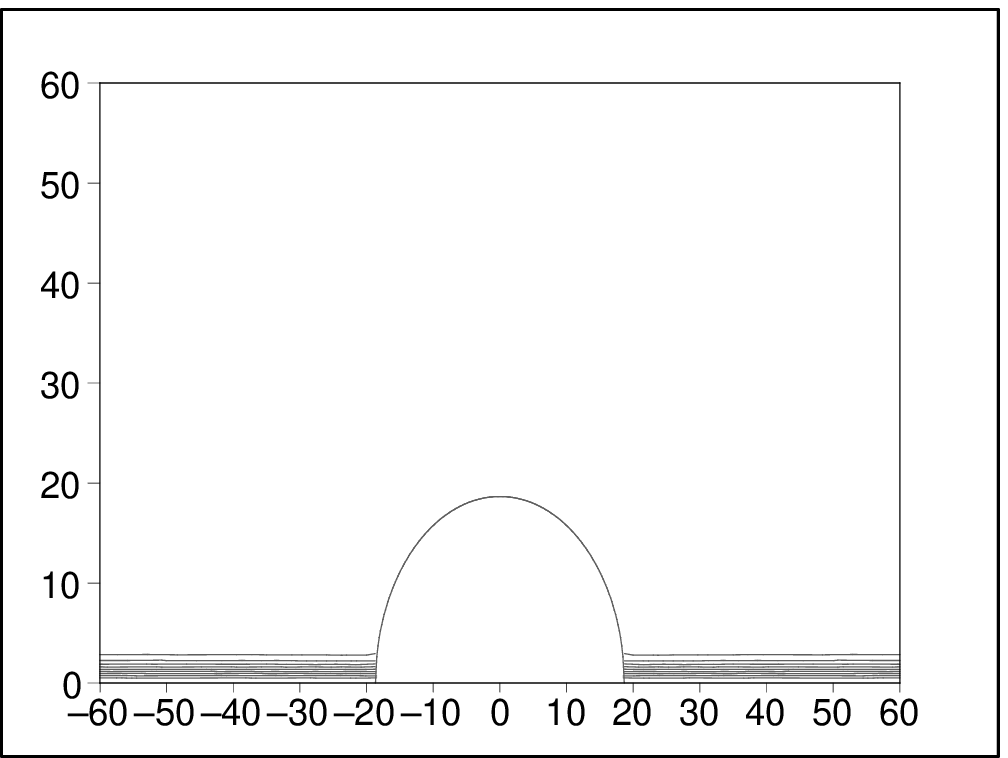,width=0.4%
\linewidth}
\end{center}
\caption{{}$X$ and $P$ contours ($X$ and $P$ increase from $0.1$ to $0.9$
upward and downward respectively) for a string in the presence of Kerr black
hole. The mass of black hole is equal to $10,$ its rotation parameter is
equal to $5$ and horizon is located in $r_{H}=18.66.$}
\label{fig9}
\end{figure}

\begin{figure}[tbp]
\begin{center}
\epsfig{file=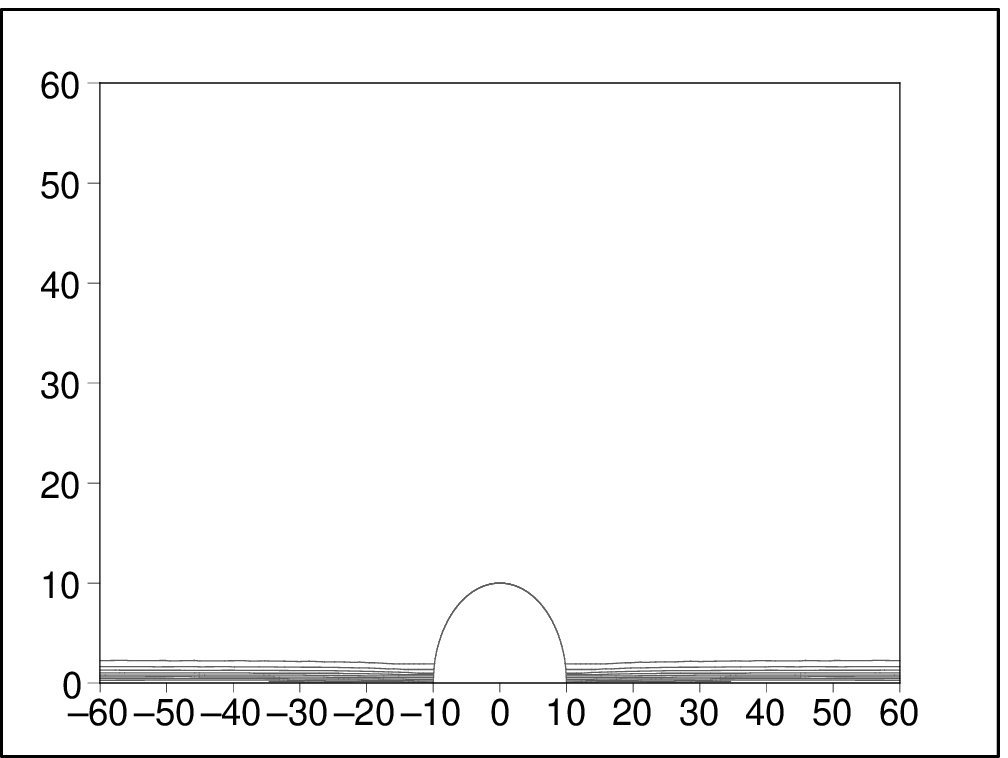,width=0.4\linewidth} %
\epsfig{file=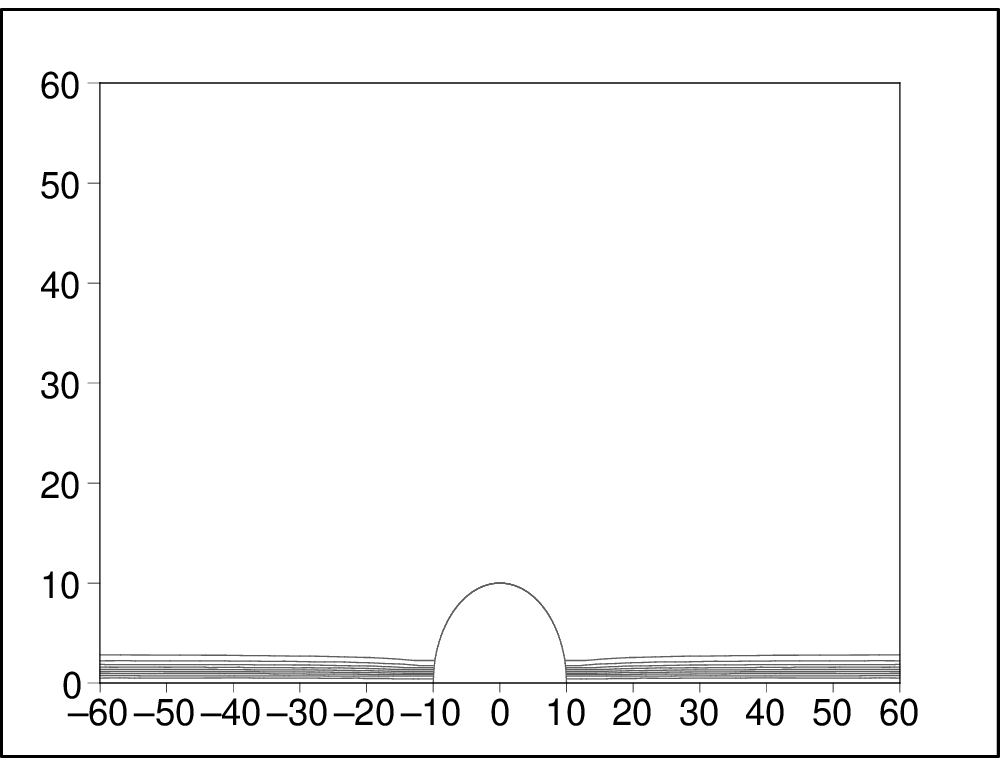,width=0.4\linewidth}
\end{center}
\caption{{}$X$ and $P$ contours ($X$ and $P$ increase from $0.1$ to $0.9$
upward and downward respectively) for a string in the presence of extremal
Kerr black hole. The mass of black hole is equal to $10,$ its rotation
parameter is equal to $10$ and horizon is located in $r_{H}=10.0.$}
\label{fig10}
\end{figure}

We consider here a special case of the preceding section, namely the Abelian
Higgs vortex in a Kerr black hole background. The Kerr metric is a special
case of \ (\ref{kerradsmetr1}) in the limit $l\rightarrow \infty .$ The
metric is given by 
\begin{equation}
ds^{2}=-\frac{\Delta }{\rho ^{2}}(dt-a\sin ^{2}\theta d\varphi )^{2}+\frac{%
\rho ^{2}}{\Delta }dr^{2}+\rho ^{2}d\theta ^{2}+\frac{\sin ^{2}\theta }{\rho
^{2}}((r^{2}+a^{2})d\varphi -adt)^{2}  \label{kerrmetr1}
\end{equation}
where $\Delta =r^{2}-2mr+a^{2}.$ The horizon is located in $r_{H}=m+\sqrt{%
m^{2}-a^{2}}.$ The equations of motion for the $X$ and $P$ fields can be
obtained from the equations (\ref{eqx}) and (\ref{eqp}) in the limit of $\
l\rightarrow \infty .$ They are, 
\begin{equation}
\begin{array}{c}
\frac{r^{2}+a^{2}-2mr}{r^{2}+a^{2}\cos ^{2}\theta }\frac{\partial
^{2}X(r,\theta )}{\partial r^{2}}+2\frac{r-m}{r^{2}+a^{2}\cos ^{2}\theta }%
\frac{\partial X(r,\theta )}{\partial r}+\frac{1}{r^{2}+a^{2}\cos ^{2}\theta 
}\frac{\partial ^{2}X(r,\theta )}{\partial \theta ^{2}}+\cot \theta \frac{1}{%
r^{2}+a^{2}\cos ^{2}\theta }\frac{\partial X(r,\theta )}{\partial \theta }
\\ 
-\frac{r^{2}+a^{2}\cos ^{2}\theta -2mr}{\sin ^{2}\theta
(r^{2}+a^{2}-2mr)(r^{2}+a^{2}\cos ^{2}\theta )}N^{2}X(r,\theta )P(r,\theta
)^{2}-\frac{1}{2}X(r,\theta )(X(r,\theta )^{2}-1)=0
\end{array}
\label{eqxkerr}
\end{equation}
\begin{equation}
\begin{array}{c}
-\frac{r^{2}+a^{2}-2mr}{r^{2}+a^{2}\cos ^{2}\theta }\frac{\partial
^{2}P(r,\theta )}{\partial r^{2}}-2m\frac{(a^{2}+r^{2})(r^{2}-a^{2}\cos
^{2}\theta )}{(r^{2}+a^{2}\cos ^{2}\theta )^{3}}\frac{\partial P(r,\theta )}{%
\partial r}-\frac{1}{r^{2}+a^{2}\cos ^{2}\theta }\frac{\partial
^{2}P(r,\theta )}{\partial \theta ^{2}} \\ 
+\cot \theta (\frac{1}{r^{2}+a^{2}\cos ^{2}\theta }+4\frac{rma^{2}\sin
^{2}\theta }{(r^{2}+a^{2}\cos ^{2}\theta )^{3}})\frac{\partial P(r,\theta )}{%
\partial \theta }+\alpha P(r,\theta )X^{2}(r,\theta )=0
\end{array}
\label{eqpkerr}
\end{equation}
We solved the differential equations (\ref{eqxkerr}) and (\ref{eqpkerr})
numerically, using the same approach employed in solving (\ref{eqx}) and (%
\ref{eqp}). Some typical results of the calculation are displayed in figures
(\ref{fig8}), (\ref{fig9}) and (\ref{fig10}). In these calculations, we
consider a string with unit winding number in the background (\ref{kerrmetr1}%
) with $a=0,5,9${\LARGE \ }along\ the extremal case{\large \ }$\ a=10$.
Throughout the black hole mass is taken to be $m=10$, and so the horizon is
respectively located at $r_{H}=20,18.66,14.36$, with $r_{H}=m=10$ in the
extremal case.

\begin{figure}[tbp]
\begin{center}
\epsfig{file=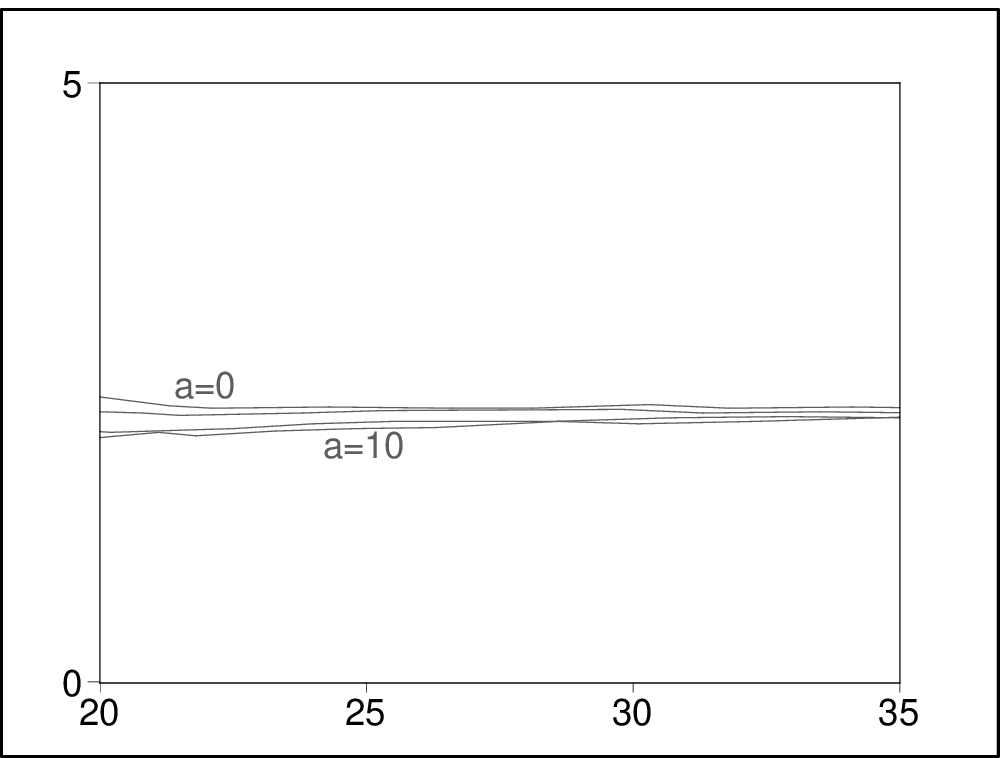,width=0.4\linewidth} %
\epsfig{file=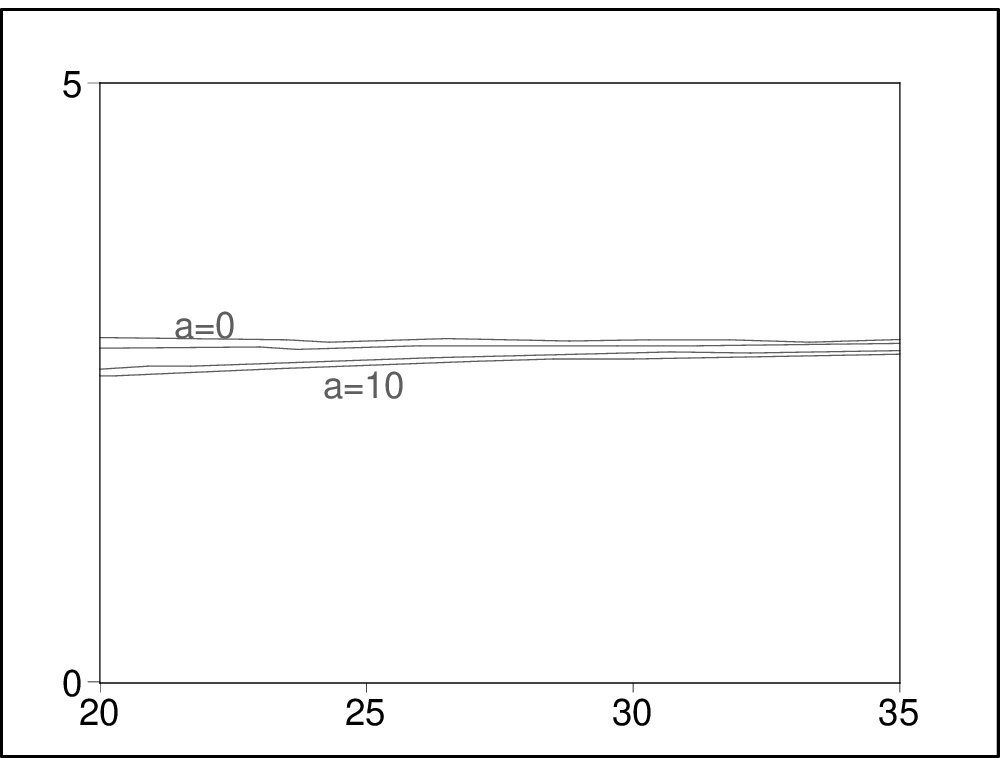,width=0.4\linewidth}
\end{center}
\caption{{}$X=0.9$ and $P=0.1$ contours in the Kerr black hole background of
mass $m=10$ for different values of the rotation parameter $a=0,5,9$ and $%
10. $ The black hole horizon is located off to the left of the figure at $%
r_{H}=20,18.66,14.36$ and $10$ respectively.}
\label{fig11}
\end{figure}

The figures in (\ref{fig8}), when $a=0$ are exactly the same as the results
of the Schwarzschild black hole \cite{Achu},\cite{Deh}.\ We note that in
changing of the rotation parameter from a value of $\ 0$ (corresponding to a
Schwarzschild black hole) to the extremal value of $10$, both the $X$ and $P$
contours undergo only a small change. In fact, as the diagrams in figure (%
\ref{fig11}) show, we observe a change on the $X=0.9$ and $P=0.1$ contours.
By increasing the rotation parameter, both the $X$ and $P$ contours decrease
a little. Similar arguments regarding the decrease of the horizon of a
Kerr-AdS black hole with increasing rotation parameter also apply to the
Kerr black hole, which explains the decrease of the string fields $X$\ and $%
P $\ in figure (\ref{fig11}) with increasing $a$. In the extremal case we
find that an extremal horizon is indeed pierced by a thin string, so the
vortex flux is not expelled from the black hole.

\section{Abelian Higgs Vortex in Reissner-Nordstrom-AdS and
Reissner-Nordstrom black holes}

In this section, we consider the Abelian Higgs vortex Lagrangian (\ref{Lag})
in the background of a charged black hole. The background metric is given
by, 
\begin{equation}
ds^{2}=-(1-\frac{2m}{r}+\frac{Q^{2}}{r^{2}}+\frac{r^{2}}{l^{2}})dt^{2}+\frac{%
1}{1-\frac{2m}{r}+\frac{Q^{2}}{r^{2}}+\frac{r^{2}}{l^{2}}}%
dr^{2}+r^{2}(d\theta ^{2}+\sin ^{2}\theta d\varphi ^{2})  \label{RNADSmetric}
\end{equation}
$Q$ is the total charge of black hole which measured by a far observer
located at $r\gg 2M,Q,$ and the black hole horizon $r_{H}$ is located at $\ $%
the largest real root of the equation $%
r^{4}+l^{2}r^{2}-2ml^{2}r+Q^{2}l^{2}=0.$ In the special case of $%
l\rightarrow \infty $, the horizon is located at $r_{H}=m+\sqrt{m^{2}-Q^{2}.}
$ The Abelian Higgs Lagrangian is as the same as (\ref{Lag}) with the real
fields $X(x^{\mu }),\omega (x^{\mu }),P_{\mu }(x^{\nu })$ again given by the
relations (\ref{XPomegadef}), whose equations of motion are (\ref{Lag2}) and
(\ref{eqmo2}). The equations of motion derived from the Lagrangian for the $%
X(r,\theta )$ and $P(r,\theta )$ fields after rescaling of coordinates in
the background (\ref{RNADSmetric}) are 
\begin{equation}
\begin{tabular}{l}
$(1-\frac{2m}{r}+\frac{Q^{2}}{r^{2}}+\frac{r^{2}}{l^{2}})\frac{\partial
^{2}X(r,\theta )}{\partial r^{2}}+\frac{2}{r}(1-\frac{m}{r}+\frac{2r^{2}}{%
l^{2}})\frac{\partial X(r,\theta )}{\partial r}+\frac{1}{r^{2}}\frac{%
\partial ^{2}X(r,\theta )}{\partial \theta ^{2}}+\frac{1}{r^{2}}\frac{%
\partial X(r,\theta )}{\partial \theta }\cot \theta -\frac{1}{2}%
(X^{3}(r,\theta )$ \\ 
$-X(r,\theta ))-N^{2}\frac{X(r,\theta )P^{2}(r,\theta )}{r^{2}\sin
^{2}\theta }=0$%
\end{tabular}
\label{eqxrnads}
\end{equation}
\begin{equation}
\begin{tabular}{l}
$(1-\frac{2m}{r}+\frac{Q^{2}}{r^{2}}+\frac{r^{2}}{l^{2}})\frac{\partial
^{2}P(r,\theta )}{\partial r^{2}}+\frac{2}{r}(\frac{m}{r}+\frac{r^{2}}{l^{2}}%
-\frac{Q^{2}}{r^{2}})\frac{\partial P(r,\theta )}{\partial r}+\frac{1}{r^{2}}%
\frac{\partial ^{2}P(r,\theta )}{\partial \theta ^{2}}-\frac{\cot \theta }{%
r^{2}}\frac{\partial P(r,\theta )}{\partial \theta }$ \\ 
$-\alpha P(r,\theta )X^{2}(r,\theta )=0$%
\end{tabular}
\label{eqprnads}
\end{equation}

We note that in the special case $Q=0$, the equations of motion (\ref
{eqxrnads}) and (\ref{eqprnads}) reduce to the equations of motion of the
vortex in the background of AdS-Schwarzschild background studied in \cite
{Deh2}. Also we note that in the special case of $\ l\rightarrow \infty ,$
the above mentioned equations reduce to the equations of motion in the
background of the Reissner-Nordstrom black hole discussed in \cite{EM1} and 
\cite{EM2}.

\begin{figure}[tbp]
\begin{center}
\epsfig{file=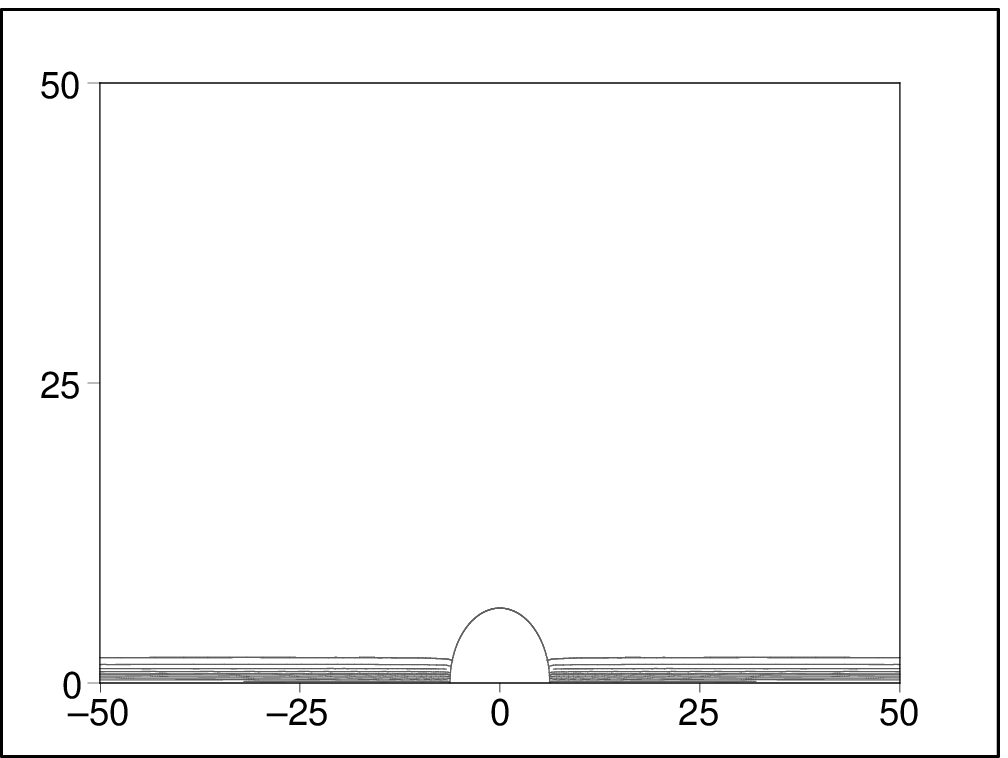,width=0.4\linewidth} %
\epsfig{file=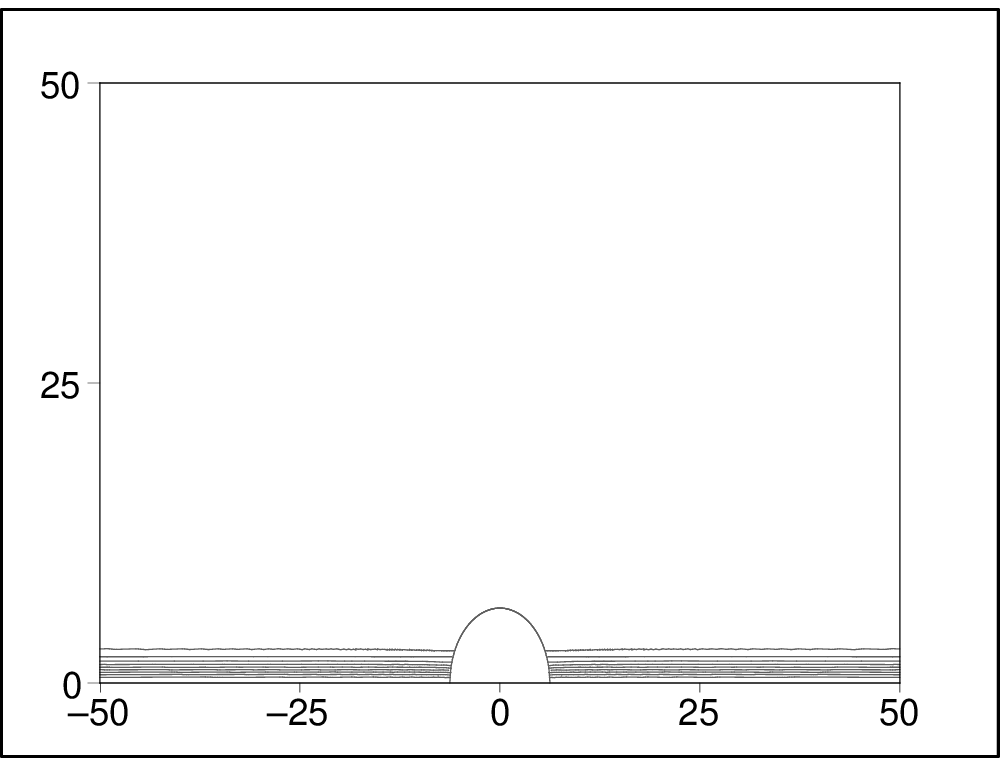,width=0.4\linewidth}
\end{center}
\caption{{}$X$ and $P$ contours ($X$ and $P$ increase from $0.1$ to $0.9$
upward and downward respectively) for a string in the presence of
Reissner-Nordstrom-AdS black hole with $l=5$. The mass of black hole is
equal to $10,$ its charge parameter is equal to $5$ and horizon is located
in $r_{H}=6.25.$}
\label{fig115}
\end{figure}

\begin{figure}[tbp]
\begin{center}
\epsfig{file=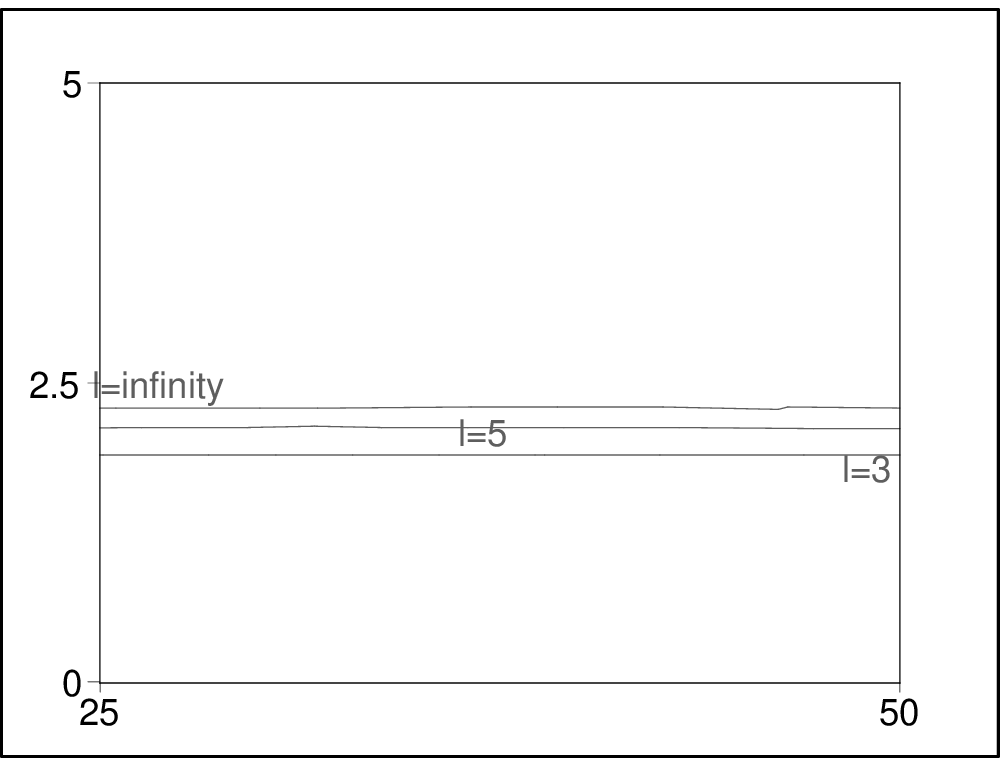,width=0.4\linewidth} %
\epsfig{file=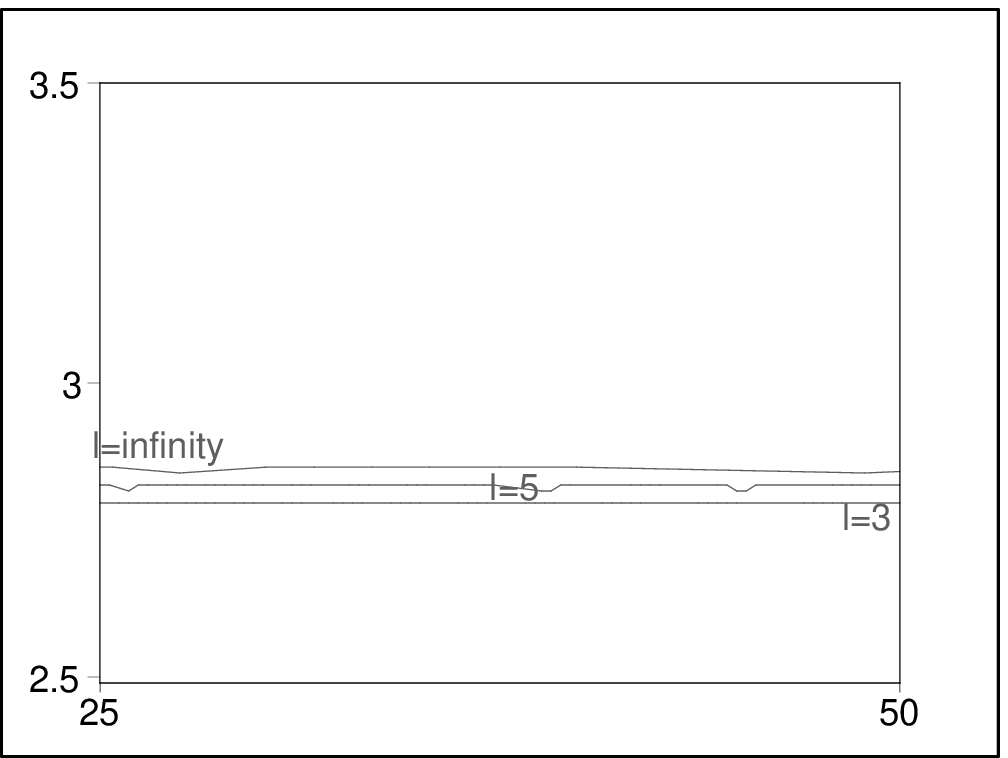,width=0.4\linewidth}
\end{center}
\caption{{}$X=0.9$ and $P=0.1$ contours in the Reissner-Nordstrom-AdS black
hole background of mass $m=10$ and charge $Q=5$ for different values of the
cosmological constant $l=3,5,\infty .$ The black hole horizon is located off
to the left of the figure at $r_{H}=4.47,6.25,18.66$ respectively.}
\label{fig116}
\end{figure}

\begin{figure}[tbp]
\begin{center}
\epsfig{file=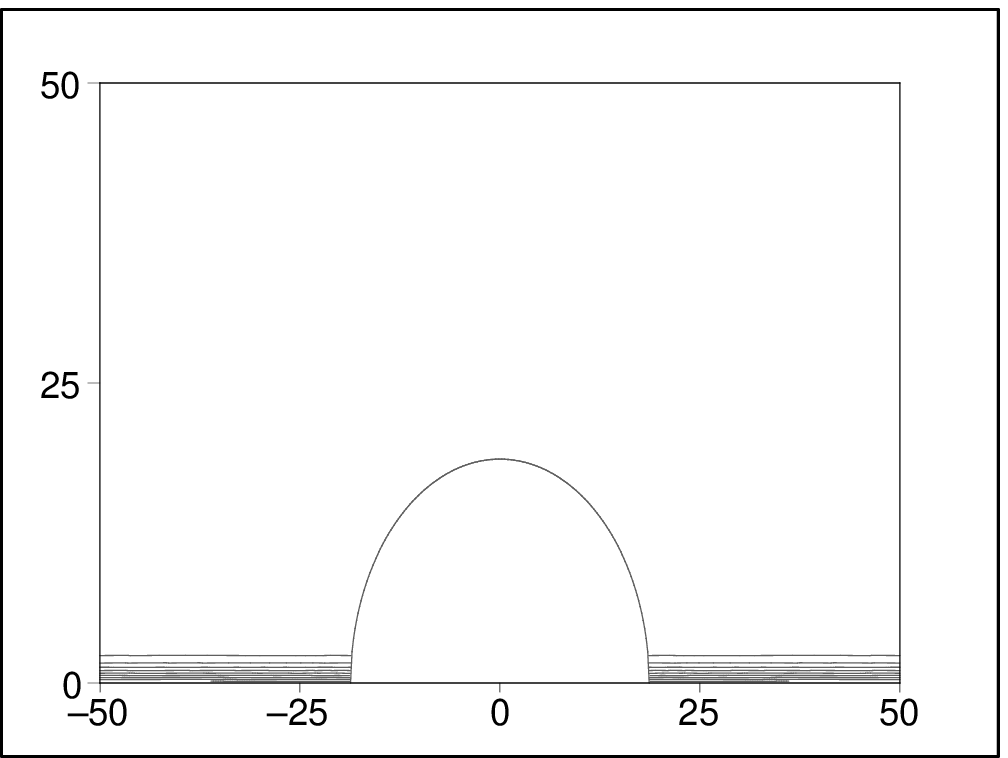,width=0.4\linewidth} \epsfig{file=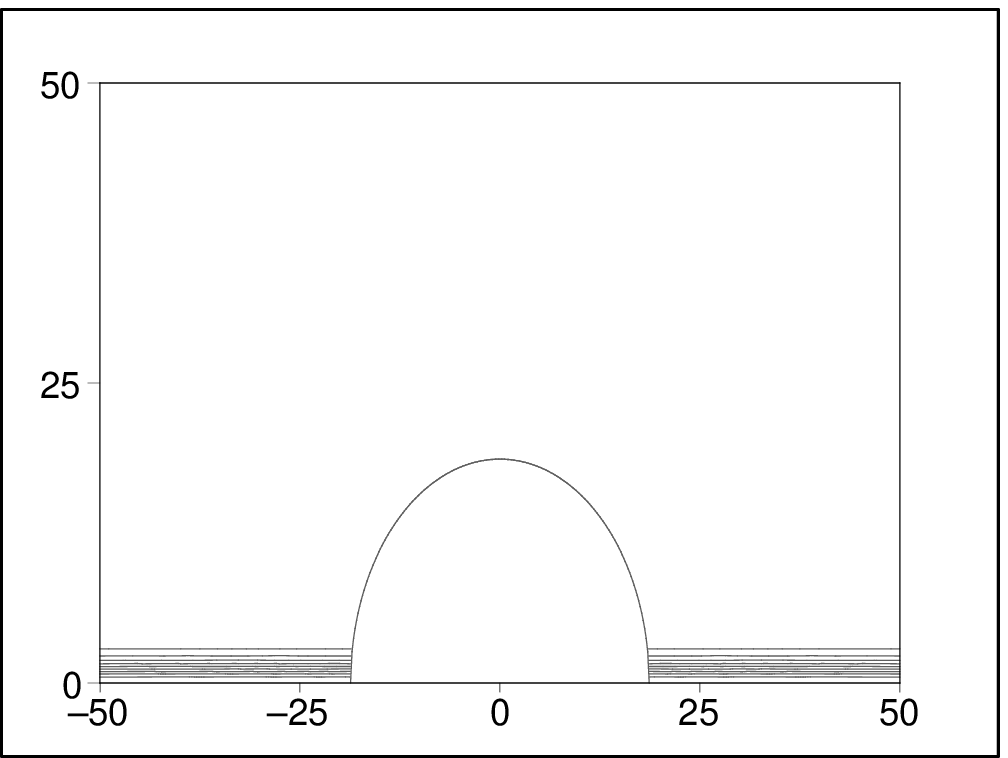,width=0.4%
\linewidth}
\end{center}
\caption{{}$X$ and $P$ contours ($X$ and $P$ increase from $0.1$ to $0.9$
upward and downward respectively) for a string in the presence of
Reissner-Nordstrom black hole. The mass of black hole is equal to $10,$ its
charge is equal to $5$ and horizon is located in $r_{H}=18.66.$}
\label{fig12}
\end{figure}

\begin{figure}[tbp]
\begin{center}
\epsfig{file=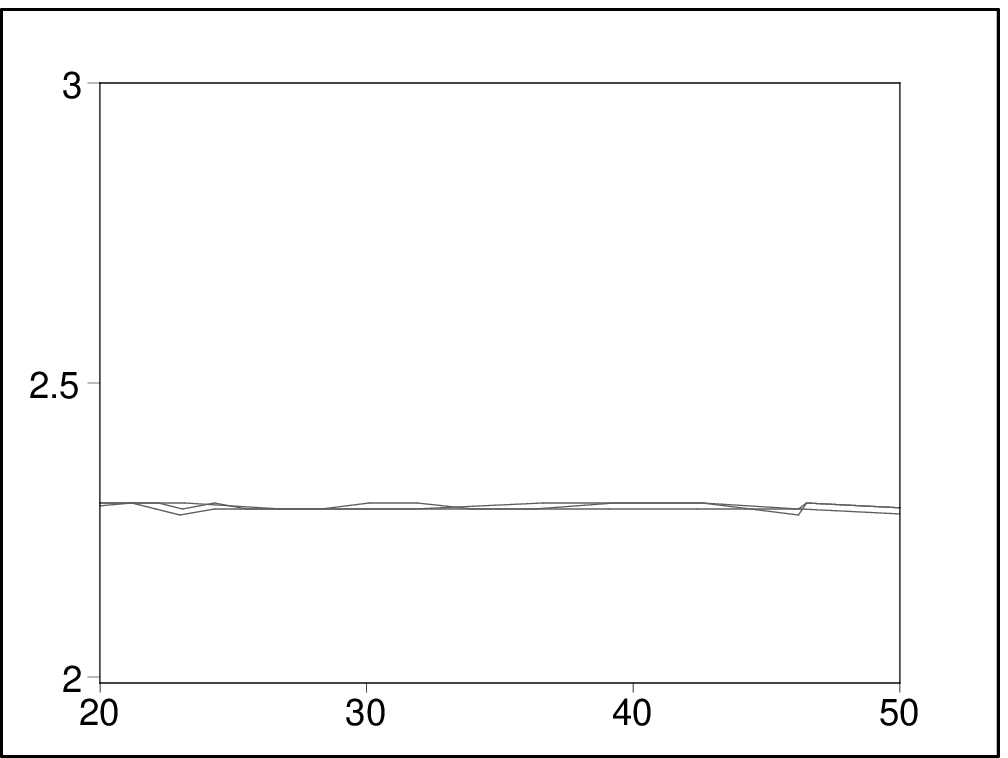,width=0.4\linewidth} %
\epsfig{file=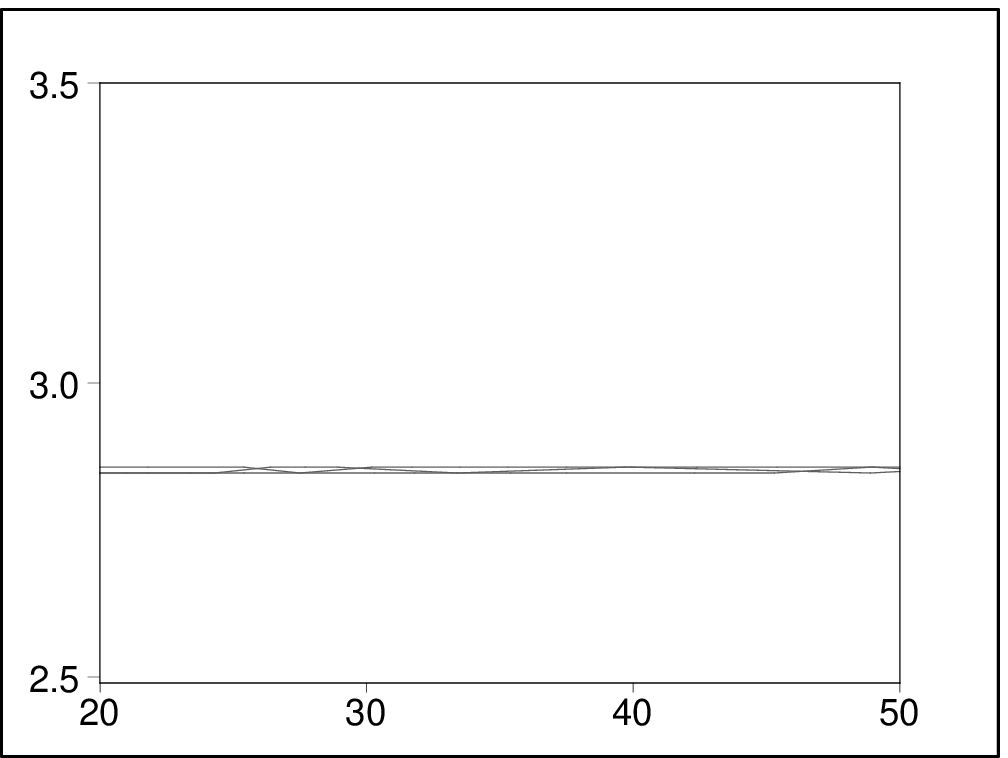,width=0.4\linewidth}
\end{center}
\caption{{}$X=0.9$ and $P=0.1$ contours in the Reissner-Nordstrom black hole
background of mass $m=10$ for different values of the charge $Q=0,5,10.$ The
black hole horizon is located off to the left of the figure at $%
r_{H}=20,18.66,10$ respectively.}
\label{fig13}
\end{figure}

We consider the static case of a string solution with winding number one.
Using the overrelaxation method described in section two, we solve
numerically the equations of motion. A typical result for the string fields
in the background of the Reissner-Nordstrom-AdS black hole with $l=5$ and $%
Q=5$ is presented in figure (\ref{fig115}). As figure (\ref{fig116}) shows,
by increasing the parameter $l$ from $3$ to $\infty ,$ the string core
changes a little. Such a tiny increase of string core also has been observed
in the case of AdS-Schwarzschild black hole by increasing the parameter $l$ 
\cite{Deh2}. In the special case of $l\rightarrow \infty $, a typical
solution of (\ref{eqxrnads}) and (\ref{eqprnads}) for a Reissner-Nordstrom
black hole with $m=10$ and $Q=5$\ is shown in figure (\ref{fig12}). In
figure (\ref{fig13}), the $X=0.9$ and $P=0.1$ contours are plotted for three
different values of the charge parameter $Q=0,5,10.$ The last case
corresponds to extremal case $Q=m=10.$ So, we observe that despite the
smaller horizon radius for a black hole with more charge, the string core
does not change drastically for the charged black holes for a wide range of
the charge parameter $0\leq Q\leq 10$. In the extremal case we find that an
extremal horizon is indeed pierced by a thin string, confirming earlier
computations carried out in references \cite{EM3} and \cite{EM4}, and that \
the vortex flux is not expelled from the black hole. In these calculations,
we have used a $600\times 700$ grid in the $(r,\theta )$ directions, a much
finer resolution than employed in previous studies \cite{EM1,EM2,EM3,EM4}.
Note that there are no values of $Q$ or $m$ for which the black hole becomes
extremal in the asymptotically AdS case.

\section{Vortex Self Gravity and Kerr Black Holes}

In this section, we study the effect of the vortex on the Kerr black hole.
The Kerr metric (\ref{kerrmetr1}) can be written in the axisymmetric Weyl
form as follows \cite{Kramer}:

\begin{equation}
ds^{2}=-e^{2U_{0}}(dt+A_{0}d\varphi )^{2}+e^{2(K_{0}-U_{0})}(d\varrho
^{2}+dz^{2})+e^{-2U_{0}}\varrho ^{2}d\varphi ^{2}  \label{kerraxis}
\end{equation}
where the functions $A_{0},U_{0},K_{0}$ are independent of $t,\varphi $ and
are given by,

\begin{equation}
\begin{array}{c}
A_{0}=am\frac{(4\sigma ^{2}-(r_{+}-r_{-})^{2})(2m+r_{+}+r_{-})}{\sigma
^{2}[(r_{+}+r_{-})^{2}-4m^{2}]+a^{2}(r_{+}-r_{-})^{2}} \\ 
U_{0}=\frac{1}{2}\ln \left\{ \frac{\sigma
^{2}[(r_{+}+r_{-})^{2}-4m^{2}]+a^{2}(r_{+}-r_{-})^{2}}{\sigma
^{2}(r_{+}+r_{-}+2m)^{2}+a^{2}(r_{+}-r_{-})^{2}}\right\} \\ 
K_{0}=\frac{1}{2}\ln \left\{ \frac{\sigma
^{2}[(r_{+}+r_{-})^{2}-4m^{2}]+a^{2}(r_{+}-r_{-})^{2}}{4\sigma ^{2}r_{+}r_{-}%
}\right\}
\end{array}
\label{axisfun}
\end{equation}
The quantities $r_{+}$ , $r_{-}$ and $\sigma $ are given by 
\begin{equation}
\begin{array}{c}
r_{\pm }=\sqrt{\varrho ^{2}+(z\pm \sigma )^{2}} \\ 
\sigma =\sqrt{m^{2}-a^{2}}
\end{array}
\label{rplusminus}
\end{equation}
The transformation between the spherical coordinates in (\ref{kerrmetr1})
and Weyl coordinates in (\ref{kerraxis}) is given by, 
\begin{equation}
\begin{array}{c}
\varrho =\sqrt{\Delta }\sin \theta =\sqrt{r^{2}-2mr+a^{2}}\sin \theta \\ 
z=(r-m)\cos \theta
\end{array}
\label{weyl}
\end{equation}
These Weyl coordinates are useful for studying the back reaction of the
vortex on the Kerr black hole.

In order to get the gravitational effect of the vortex on the Kerr black
hole, we consider a general static axisymmetric metric of the form 
\begin{equation}
ds^{2}=-e^{2U}(dt+Ad\varphi )^{2}+e^{2(K-U)}(d\varrho
^{2}+dz^{2})+e^{-2U}\xi ^{2}d\varphi ^{2}  \label{genaxis}
\end{equation}
where $A,U,K$ and $\xi $ are dependent on the Weyl coordinates $\varrho $
and $z.$ The Einstein equations become 
\begin{equation}
\frac{\partial ^{2}\xi }{\partial \varrho ^{2}}+\frac{\partial ^{2}\xi }{%
\partial z^{2}}=\varepsilon (T_{z}^{z}+T_{\varrho }^{\varrho })\sqrt{-g}
\label{eineqs1}
\end{equation}
\begin{equation}
\begin{array}{c}
e^{4U}\{(\frac{\partial A}{\partial \varrho })^{2}+(\frac{\partial A}{%
\partial z})^{2}\}\xi +2(\frac{\partial ^{2}U}{\partial \varrho ^{2}}+\frac{%
\partial ^{2}U}{\partial z^{2}})\xi \\ 
+2\frac{\partial U}{\partial z}\frac{\partial \xi }{\partial z}+2\frac{%
\partial U}{\partial \varrho }\frac{\partial \xi }{\partial \varrho }%
+e^{4U}A\xi (\frac{\partial ^{2}A}{\partial \varrho ^{2}}+\frac{\partial
^{2}A}{\partial z^{2}}) \\ 
+3e^{4U}A(\frac{\partial A}{\partial \varrho }\frac{\partial \xi }{\partial
\varrho }+\frac{\partial A}{\partial z}\frac{\partial \xi }{\partial z}%
)+4e^{4U}\{A\xi (\frac{\partial A}{\partial \varrho }\frac{\partial U}{%
\partial \varrho }+\frac{\partial A}{\partial z}\frac{\partial U}{\partial z}%
) \\ 
+A^{2}(\frac{\partial A}{\partial \varrho }\frac{\partial \xi }{\partial
\varrho }+\frac{\partial A}{\partial z}\frac{\partial \xi }{\partial z}%
)\}+e^{4U}A^{2}(\frac{\partial ^{2}\xi }{\partial \varrho ^{2}}+\frac{%
\partial ^{2}\xi }{\partial z^{2}})=-\varepsilon (T_{t}^{t}-T_{\varphi
}^{\varphi }+T_{z}^{z}+T_{\varrho }^{\varrho })\sqrt{-g}
\end{array}
\label{eineqs2}
\end{equation}
\begin{equation}
\begin{array}{c}
e^{4U}\frac{\partial A}{\partial \varrho }\frac{\partial A}{\partial z}\xi
^{2}+e^{4U}\frac{\partial \xi }{\partial \varrho }\frac{\partial \xi }{%
\partial z}A^{2}+e^{4U}A\xi (\frac{\partial A}{\partial \varrho }\frac{%
\partial \xi }{\partial z}+\frac{\partial A}{\partial z}\frac{\partial \xi }{%
\partial \varrho }) \\ 
-2\xi \frac{\partial ^{2}\xi }{\partial \varrho \partial z}-4\xi ^{2}\frac{%
\partial U}{\partial \varrho }\frac{\partial U}{\partial z}+2\xi (\frac{%
\partial \xi }{\partial \varrho }\frac{\partial K}{\partial z}+\frac{%
\partial \xi }{\partial z}\frac{\partial K}{\partial \varrho })=2\varepsilon
\xi (T_{z}^{\varrho })\sqrt{-g}
\end{array}
\label{eineqs3}
\end{equation}
\begin{equation}
\begin{array}{c}
4\xi ^{2}\{(\frac{\partial U}{\partial \varrho })^{2}+(\frac{\partial U}{%
\partial z})^{2}\}+4e^{4U}(\frac{\partial A}{\partial \varrho }\frac{%
\partial \xi }{\partial \varrho }+\frac{\partial A}{\partial z}\frac{%
\partial \xi }{\partial z})A\xi -e^{4U}\{(\frac{\partial \xi }{\partial
\varrho })^{2}+(\frac{\partial \xi }{\partial z})^{2}\}A^{2} \\ 
+4(\frac{\partial ^{2}K}{\partial \varrho ^{2}}+\frac{\partial ^{2}K}{%
\partial z^{2}})\xi ^{2}+8e^{4U}(\frac{\partial A}{\partial \varrho }\frac{%
\partial U}{\partial \varrho }+\frac{\partial A}{\partial z}\frac{\partial U%
}{\partial z})A\xi ^{2} \\ 
+2e^{4U}(\frac{\partial ^{2}A}{\partial \varrho ^{2}}+\frac{\partial ^{2}A}{%
\partial z^{2}})A\xi ^{2}+8e^{4U}\xi A^{2}(\frac{\partial U}{\partial
\varrho }\frac{\partial \xi }{\partial \varrho }+\frac{\partial U}{\partial z%
}\frac{\partial \xi }{\partial z})+2e^{4U}\xi A^{2}(\frac{\partial ^{2}\xi }{%
\partial \varrho ^{2}}+\frac{\partial ^{2}\xi }{\partial z^{2}}) \\ 
+e^{4U}\xi ^{2}\{(\frac{\partial A}{\partial \varrho })^{2}+(\frac{\partial A%
}{\partial z})^{2}\} \\ 
=4\varepsilon \xi (T_{\varphi }^{\varphi })\sqrt{-g}
\end{array}
\label{eineqs4}
\end{equation}
where $\varepsilon =8\pi \eta ^{2}$ and $g$ is the determinant of the metric
(\ref{genaxis}). The non-vanishing energy-momentum components are given by 
\begin{equation}
\begin{array}{c}
T_{t}^{t}=2(X^{2}-1)^{2}-\frac{e^{2U}X^{2}P^{2}}{\xi ^{2}}-e^{2(U-K)}\left\{
(\frac{\partial X}{\partial \varrho })^{2}+(\frac{\partial X}{\partial z}%
)^{2}\right\} +e^{2(2U-K)}\frac{1}{\xi ^{2}}\left\{ (\frac{\partial P}{%
\partial \varrho })^{2}+(\frac{\partial P}{\partial z})^{2}\right\} \\ 
T_{\varrho }^{\varrho }=2(X^{2}-1)^{2}-\frac{e^{2U}X^{2}P^{2}}{\xi ^{2}}%
+e^{2(U-K)}\left\{ (\frac{\partial X}{\partial \varrho })^{2}-(\frac{%
\partial X}{\partial z})^{2}\right\} -e^{2(2U-K)}\frac{1}{\xi ^{2}}\left\{ (%
\frac{\partial P}{\partial \varrho })^{2}-(\frac{\partial P}{\partial z}%
)^{2}\right\} \\ 
T_{z}^{z}=2(X^{2}-1)^{2}-\frac{e^{2U}X^{2}P^{2}}{\xi ^{2}}-e^{2(U-K)}\left\{
(\frac{\partial X}{\partial \varrho })^{2}-(\frac{\partial X}{\partial z}%
)^{2}\right\} +e^{2(2U-K)}\frac{1}{\xi ^{2}}\left\{ (\frac{\partial P}{%
\partial \varrho })^{2}-(\frac{\partial P}{\partial z})^{2}\right\} \\ 
T_{\varphi }^{\varphi }=2(X^{2}-1)^{2}+\frac{e^{2U}X^{2}P^{2}}{\xi ^{2}}%
-e^{2(U-K)}\left\{ (\frac{\partial X}{\partial \varrho })^{2}+(\frac{%
\partial X}{\partial z})^{2}\right\} -e^{2(2U-K)}\frac{1}{\xi ^{2}}\left\{ (%
\frac{\partial P}{\partial \varrho })^{2}+(\frac{\partial P}{\partial z}%
)^{2}\right\} \\ 
T_{z}^{\varrho }=2e^{2(U-K)}\frac{\partial X}{\partial \varrho }\frac{%
\partial X}{\partial z}-2e^{2(2U-K)}\frac{1}{\xi ^{2}}\frac{\partial P}{%
\partial \varrho }\frac{\partial P}{\partial z}
\end{array}
\label{emt}
\end{equation}

Writing $U=U_{0}+\varepsilon U_{1},$ $K=K_{0}+\varepsilon
K_{1},A=A_{0}+\varepsilon A_{1},\xi =\xi _{0}+\varepsilon \xi _{1},$ where $%
\xi _{0}=\varrho ,$ we can solve the equations (\ref{eineqs1})-(\ref{eineqs4}%
) to first order in $\varepsilon .$ To order zero in $\varepsilon ,$ the
background metric is give by (\ref{kerraxis}) and the string fields by $%
X_{0}(R)$ and $P_{0}(R).$ So to first order in $\varepsilon ,$ equation (\ref
{eineqs1}) becomes 
\begin{equation}
\frac{\partial ^{2}\xi _{1}}{\partial \varrho ^{2}}+\frac{\partial ^{2}\xi
_{1}}{\partial z^{2}}=e^{2(K_{0}-U_{0})}\varrho (T_{z}^{z}+T_{\varrho
}^{\varrho })_{0}  \label{feineqs}
\end{equation}
which can be solved by assuming the form $\xi _{1}=\varrho \zeta (R).$ Here,
we take $R=\varrho e^{-U_{0}}.$ Then function $\zeta (R)$ satisfies 
\begin{equation}
\frac{d^{2}\zeta }{dR^{2}}+\frac{2}{R}\frac{d\zeta }{dR}=(T_{z}^{z}+T_{%
\varrho }^{\varrho })_{0}  \label{zeta}
\end{equation}
which can be solved to give 
\begin{equation}
\zeta (R)=\int \frac{\int R^{2}(T_{z}^{z}+T_{\varrho }^{\varrho })_{0}dR}{%
R^{2}}dR=\int R(T_{z}^{z}+T_{\varrho }^{\varrho })_{0}dR-\frac{1}{R}\int
R^{2}(T_{z}^{z}+T_{\varrho }^{\varrho })_{0}dR  \label{zetasol}
\end{equation}

Since outside the vortex core the string fields $X_{0}(R)$ and $P_{0}(R)$
rapidly approach constant values, the integrals in (\ref{zetasol}) go
correspondingly to the respective constant values $\tau _{1\text{ }}$and $%
\tau _{2}.$ This gives $\xi =\rho (1+\varepsilon \tau _{1})$, and the
solutions of the other coupled equations (\ref{eineqs2})-(\ref{eineqs4}) are
found to be $U=U_{0}+\varepsilon \upsilon ,$ \ $K=K_{0}+2\varepsilon
\upsilon $ and $A=A_{0}(1+\varepsilon \tau _{1})e^{-2\varepsilon \upsilon }$ 
$\ $where $\upsilon =\frac{1}{2}\int \frac{\int R(T_{t}^{t}-T_{\varphi
}^{\varphi }+T_{z}^{z}+T_{\varrho }^{\varrho })_{0}dR}{R}dR.$ Using these
quantities in (\ref{genaxis}) and rescaling the coordinates $t,\varrho ,z$
and function $A_{0}$, by $\widehat{t}=e^{\varepsilon \upsilon }t,\widehat{%
\varrho }=e^{\varepsilon \upsilon }\varrho ,\widehat{z}=e^{\varepsilon
\upsilon }z$ and $\widehat{A}_{0}=e^{\varepsilon \upsilon }A_{0},$ we get
finally, 
\begin{equation}
ds^{2}=-e^{2U_{0}}\{d\widehat{t}+\widehat{A}_{0}(1+\varepsilon \tau
_{1})e^{-2\varepsilon \upsilon }d\varphi \}^{2}+e^{2(K_{0}-U_{0})}(d\widehat{%
\varrho }^{2}+d\widehat{z}^{2})+e^{-2U_{0}}\widehat{\varrho }%
^{2}(1+\varepsilon \tau _{1})^{2}e^{-4\varepsilon \upsilon }d\varphi ^{2}
\label{kerrcorr}
\end{equation}
which describes the Kerr metric with a deficit angle, since the angle $%
\widetilde{\varphi }\equiv (1+\varepsilon \tau _{1})e^{-2\varepsilon
\upsilon }\varphi $ belongs to the interval $[0,2\pi \{1+\varepsilon (\tau
_{1}-2\upsilon )\}].$

\section{Vortex Self Gravity on the Kerr-AdS and Reissner-Nordstrom-AdS
Black Holes}

We now first consider the effect of \ the vortex on the Kerr-AdS black hole.
As we have seen in \cite{Deh2}, this is a formidable problem even for the
simpler cases of \ the effect of the vortex on the AdS-Schwarzschild or
Schwarzschild black hole backgrounds.

For the AdS-Schwarzschild black hole, it has been shown that the components
of the energy-momentum tensor rapidly go to zero outside the core string,
leading to a situation similar to that of pure AdS spacetime. A full study
of vortex self gravity in pure AdS spacetime was carried out in ref. \cite
{Deh2}. We assume for the present case that the thickness of the vortex is
much smaller than all other relevant length scales and that the
gravitational effects of the string are weak enough so that the linearized
Einstein-Abelian Higgs differential equations are applicable. So, we
consider a thin string \ with the winding number $N=1$ in the Kerr-AdS
background with $l=5$. The analysis for other values of $l$ is similar. The
rescaled diagonal components of the energy-momentum tensor are 
\begin{equation}
\begin{tabular}{l}
$T_{\mu \mu }=f_{\mu \mu }^{(r)}(\frac{\partial X}{\partial r})^{2}+g_{\mu
\mu }^{(r)}(\frac{\partial P}{\partial r})^{2}+f_{\mu \mu }^{(\theta )}(%
\frac{\partial X}{\partial \theta })^{2}+g_{\mu \mu }^{(\theta )}(\frac{%
\partial P}{\partial \theta })^{2}$ \\ 
$\ \ \ \ +h_{1\mu \mu }(X^{2}-1)^{2}+h_{2\mu \mu }X^{2}P^{2}$%
\end{tabular}
\label{stress}
\end{equation}
where the functions $f_{\mu \mu }^{(r)},f_{\mu \mu }^{(\theta )},g_{\mu \mu
}^{(r)},g_{\mu \mu }^{(\theta )},h_{1\mu \mu }$ and $h_{2\mu \mu }$ are
complicated functions of \ the coordinates $\left( r,\theta \right) .$ Their
functional forms are presented in the appendix. In the figure (\ref{fig14})
the behaviour of energy-momentum tensor components for a fixed value of $\ z$
is shown. We have checked that the behaviour of the components other $z$
directions is similar.

It is clear from these figures that the components of the energy-momentum
tensor rapidly go to zero outside the core of the vortex, rendering the
situation similar to that of AdS-Schwarzschild spacetime. Performing the
same calculation as for pure AdS spacetime described in detail in \cite{Deh2}%
, we obtain the following metric for the Kerr-AdS spacetime incorporating
the effect of the vortex

\begin{equation}
ds^{2}=-\frac{\Delta _{r}}{\rho ^{2}}(dt-\frac{a\beta }{\Sigma }\sin
^{2}\theta d\varphi )^{2}+\frac{\rho ^{2}}{\Delta _{r}}dr^{2}+\frac{\rho ^{2}%
}{\Delta _{\theta }}d\theta ^{2}+\frac{\Delta _{\theta }\sin ^{2}\theta }{%
\rho ^{2}}(adt-\beta \frac{r^{2}+a^{2}}{\Sigma }d\varphi )^{2}
\label{ADSSCHDEF}
\end{equation}
which $\beta $ is a constant dependent on the different parameters of the
black hole. The above metric describes a Kerr-AdS metric with a deficit
angle. Also if we take the limiting case of $l\rightarrow \infty ,$ we get
the following Kerr spacetime incorporated the effect of the string on it, 
\[
ds^{2}=-\frac{\Delta }{\rho ^{2}}(dt-a\gamma \sin ^{2}\theta d\varphi )^{2}+%
\frac{\rho ^{2}}{\Delta }dr^{2}+\rho ^{2}d\theta ^{2}+\frac{\sin ^{2}\theta 
}{\rho ^{2}}(\gamma (r^{2}+a^{2})d\varphi -adt)^{2} 
\]
in which $\gamma $ is another constant that also depends on the different
parameters of the Kerr black hole as noted above in (\ref{kerrcorr}).

So, using a physical Lagrangian based model, we have established that the
presence of the cosmic string induces a deficit angle in the Kerr-AdS and
Kerr black holes metric. In the case of charged Reissner-Nordstrom-AdS black
hole, the energy-momentum tensor also goes rapidly to zero outside the core
string, so the above arguments are still applicable: the effect of the
vortex on the background (\ref{RNADSmetric}) simply multiplies the angle
coordinate $\varphi $ by a constant, inducing a deficit angle in the
Reissner-Nordstrom-AdS black hole spacetime.

\section{Conclusion}

The effect of a vortex on pure AdS spacetime is to create a deficit angle in
the metric in the thin vortex approximation. We have extended this result to
the charged and stationary cases, establishing numerically that Abelian
Higgs vortices of finite thickness can pierce the Kerr, Kerr-AdS, and
Reissner-Nordstrom-AdS black hole horizons. These solutions could thus be
interpreted as stable Abelian hair for these black holes.

We have obtained numerical solutions for various cosmological constants and
rotation parameter for a string with winding number one. Our solutions in
the limit \ $a\rightarrow 0$ coincide with the known solutions in the
AdS-Schwarzschild spacetime. We found that by increasing the value of
rotation parameter, the string core decreases. Inclusion of the self gravity
of the vortex in the Kerr-AdS background metric was shown to induce a
deficit angle in this metric. We extended these results to obtained
numerical solutions for a string with winding number one in the background
of a Reissner-Nordstrom-AdS black hole for different values of the
parameters.

We also considered extremal black holes in all three situations. We found by
numerical integration that the string core remains nearly the same as that
of \ a Schwarzschild black hole with the same mass. \ We observed that in
the extremal case the string can pierce the charged/rotating black hole
horizon and no flux expulsion occurs, an effect noted for charged black
holes in the asymptotically flat case \cite{EM3,EM4}. Our results establish
that flux-expulsion does not take place for extremal black holes, and that
generically a cosmic vortex will pierce both rotating and charged black
holes, with the stress-energy of the vortex inducing a deficit angle in
these black hole metrics.

Other interesting issues concern the development of \ a holographic
description of the vortex solution in these spacetimes. \ The holographic
description of a vortex in was given by a discontinuity in the logarithmic
derivative of a scalar-two point function in a CFT formulated on a spacetime
with deficit angle. \ It is reasonable to conjecture that our extensions of
vortex solutions to charged and rotating cases will yield a similar effect,
but with the CFT at a finite temperature due to the presence of the black
hole event horizon. Work on these problems is in progress.

\section{Appendix}

Here we present the coefficients $A_{ij},...,F_{ij\text{ }}^{\prime }$
appear in the equations (\ref{findiffxx}) and (\ref{findiffxp}). Let us
rename the coefficients of the $\frac{\partial ^{2}X(r,\theta )}{\partial
r^{2}},\frac{\partial X(r,\theta )}{\partial r},\frac{\partial
^{2}X(r,\theta )}{\partial \theta ^{2}},\frac{\partial X(r,\theta )}{%
\partial \theta }$ and $X(r,\theta )$ (fifth term) appearing in the equation
(\ref{eqx}) by $X_{rr},X_{r},X_{\theta \theta },X_{\theta },\widetilde{X}$
respectively. Also, we rename the coefficients of the $\frac{\partial
^{2}P(r,\theta )}{\partial r^{2}},\frac{\partial P(r,\theta )}{\partial r},%
\frac{\partial ^{2}P(r,\theta )}{\partial \theta ^{2}},\frac{\partial
P(r,\theta )}{\partial \theta }$ appearing in the equation (\ref{eqp}) by $%
P_{rr},P_{r},P_{\theta \theta },P_{\theta }$ respectively. Then the
coefficients $A_{ij},...,F_{ij}$ inside the grid points are given by the
following relations, 
\[
\begin{array}{c}
A_{ij}=\left\{ \frac{X_{rr}}{(\Delta r)^{2}}+\frac{X_{r}}{2\Delta r}\right\}
_{r=r_{i},\theta =\theta _{j}} \\ 
B_{ij}=\left\{ \frac{X_{rr}}{(\Delta r)^{2}}-\frac{X_{r}}{2\Delta r}\right\}
_{r=r_{i},\theta =\theta _{j}} \\ 
C_{ij}=\left\{ \frac{X_{\theta \theta }}{(\Delta \theta )^{2}}+\frac{%
X_{\theta }}{2\Delta \theta }\right\} _{r=r_{i},\theta =\theta _{j}} \\ 
D_{ij}=\left\{ \frac{X_{\theta \theta }}{(\Delta \theta )^{2}}-\frac{%
X_{\theta }}{2\Delta \theta }\right\} _{r=r_{i},\theta =\theta _{j}} \\ 
E_{ij}=\left\{ -2(\frac{X_{rr}}{(\Delta r)^{2}}+\frac{X_{\theta \theta }}{%
(\Delta \theta )^{2}})+\widetilde{X}\right\} _{r=r_{i},\theta =\theta _{j}}
\\ 
F_{ij}=-\frac{1}{2}X_{ij}(X_{ij}^{2}-1)
\end{array}
\]
The coefficients $A_{ij}^{\prime },...,D_{ij}^{\prime }$ have the similar
form as $A_{ij},...,D_{ij}$ with the replacements $X_{rr},X_{r},X_{\theta
\theta },X_{\theta }$ by $P_{rr},P_{r},P_{\theta \theta },P_{\theta }$ and 
\[
E_{ij}^{\prime }=-2\left\{ \frac{X_{rr}}{(\Delta r)^{2}}+\frac{X_{\theta
\theta }}{(\Delta \theta )^{2}}\right\} _{r=r_{i},\theta =\theta
_{j}}+\alpha X_{ij}^{2} 
\]
The coefficient $F_{ij}^{\prime }$ is equal to zero.

Here, we present also some of the functional form of the functions $f_{\mu
\mu }^{(r)},f_{\mu \mu }^{(\theta )},g_{\mu \mu }^{(r)},g_{\mu \mu
}^{(\theta )},h_{1\mu \mu }$ and $h_{2\mu \mu }$ which was appeared in the
formula (\ref{stress}).

For example, we have

\[
\begin{array}{c}
f_{tt}^{(r)}=\frac{%
(r^{2}l^{2}+r^{4}+a^{2}l^{2}+r^{2}a^{2}-2mrl^{2})(r^{2}l^{2}+r^{4}+r^{2}a^{2}-2mrl^{2}+a^{2}l^{2}\cos ^{2}\theta +a^{4}\cos ^{2}\theta -a^{4}\cos ^{4}\theta )%
}{2l^{4}(a^{2}\cos ^{2}\theta +r^{2})^{2}} \\ 
f_{tt}^{(\theta )}=\frac{(a^{2}\cos ^{2}\theta
-l^{2})(-r^{2}l^{2}-r^{4}-r^{2}a^{2}+2mrl^{2}-a^{2}l^{2}\cos ^{2}\theta
-a^{4}\cos ^{2}\theta +a^{4}\cos ^{4}\theta )}{2l^{4}(a^{2}\cos ^{2}\theta
+r^{2})^{2}} \\ 
g_{tt}^{(r)}=\frac{(r^{2}l^{2}+r^{4}+r^{2}a^{2}-2mrl^{2}+a^{2}l^{2}\cos
^{2}\theta +a^{4}\cos ^{2}\theta -a^{4}\cos ^{4}\theta )^{2}(a^{2}-l^{2})^{2}%
}{2\alpha l^{6}(a^{2}\cos ^{2}\theta +r^{2})^{3}\sin ^{2}\theta
(l^{2}-a^{2}\cos ^{2}\theta )} \\ 
g_{tt}^{(\theta )}=\frac{(r^{2}l^{2}+r^{4}+r^{2}a^{2}-2mrl^{2}+a^{2}l^{2}%
\cos ^{2}\theta +a^{4}\cos ^{2}\theta -a^{4}\cos ^{4}\theta
)^{2}(a^{2}-l^{2})^{2}}{2\alpha l^{6}(a^{2}\cos ^{2}\theta +r^{2})^{3}\sin
^{2}\theta (r^{2}l^{2}+r^{4}+a^{2}l^{2}+r^{2}a^{2}-2mrl^{2})} \\ 
h_{1tt}=\frac{(r^{2}l^{2}+r^{4}+r^{2}a^{2}-2mrl^{2}+a^{2}l^{2}\cos
^{2}\theta +a^{4}\cos ^{2}\theta \sin ^{2}\theta )}{l^{2}(a^{2}\cos
^{2}\theta +r^{2})} \\ 
h_{2tt}=\frac{(r^{2}l^{2}+r^{4}+r^{2}a^{2}-2mrl^{2}+a^{2}l^{2}\cos
^{2}\theta +a^{4}\cos ^{2}\theta \sin ^{2}\theta )(a^{2}-l^{2})^{2}}{%
2l^{4}(a^{2}\cos ^{2}\theta -l^{2})\sin ^{2}\theta (a^{2}\cos ^{2}\theta
+r^{2})^{2}(r^{2}l^{2}+r^{4}+a^{2}l^{2}+r^{2}a^{2}-2mrl^{2})}
\end{array}
\]

Other functions have similar complicated structures; we shall not present
them here.

\bigskip

{\Large Acknowledgments}

This work was supported by the Natural Sciences and Engineering Research
Council of Canada.

\begin{figure}[tbp]
\begin{center}
\epsfig{file=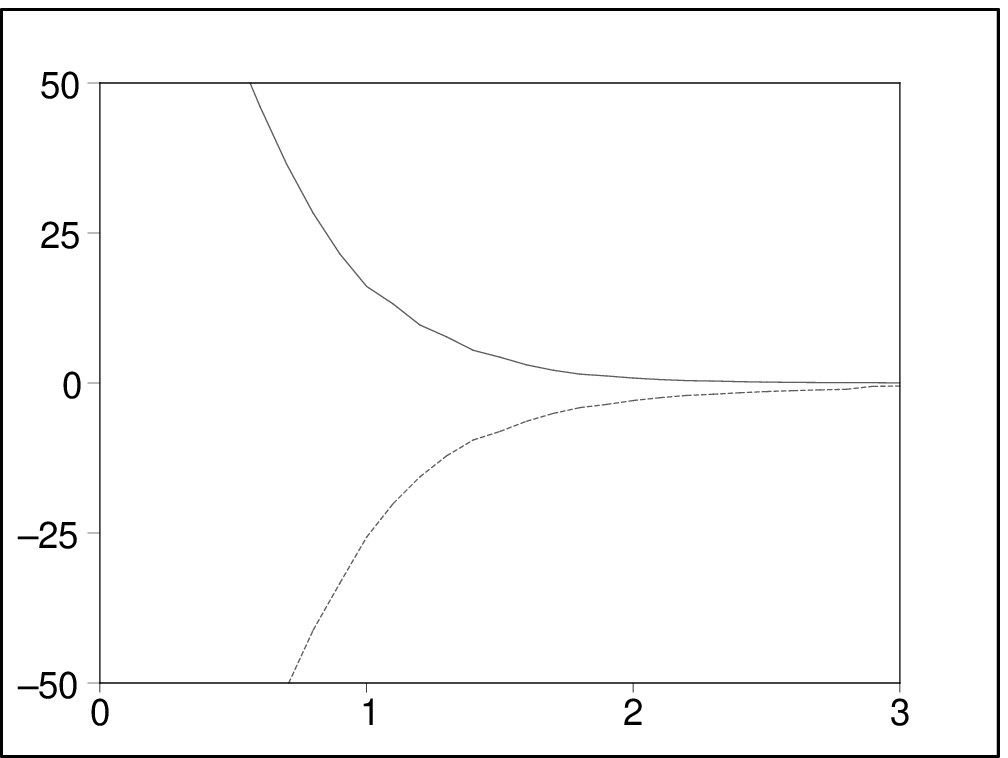,width=0.4\linewidth} %
\epsfig{file=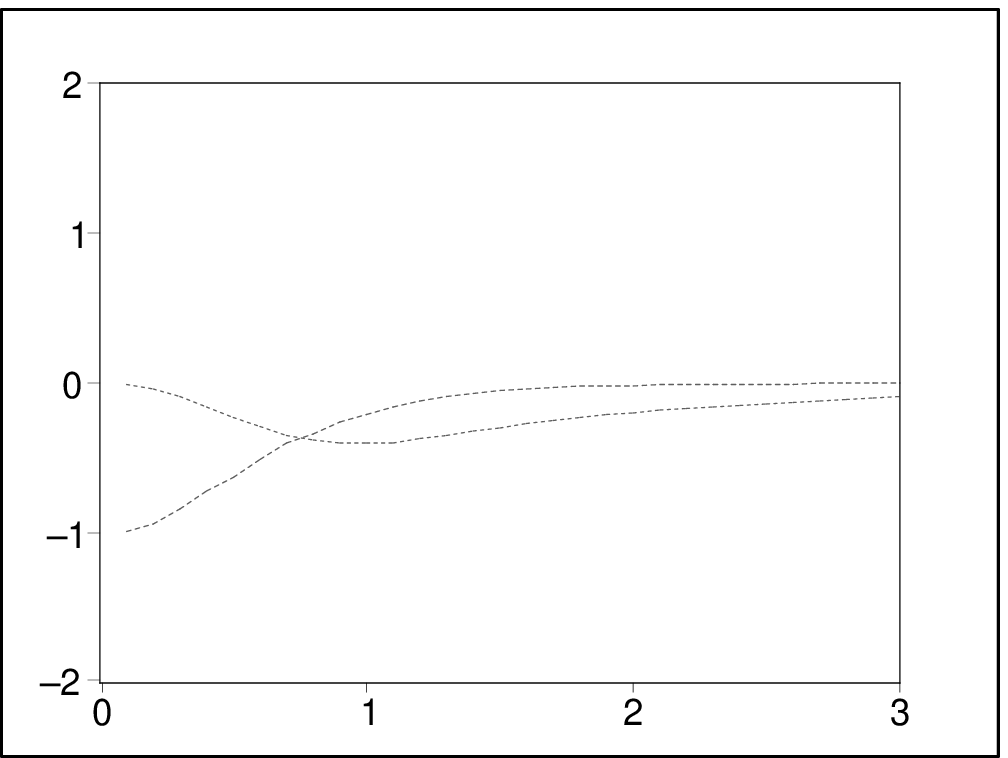,width=0.4\linewidth}
\end{center}
\caption{{}$T_{tt}$ (solid), $T_{\protect\theta \protect\theta }$ (dashed), $%
T_{\protect\varphi \protect\varphi }$ (dotted) and $T_{rr}$ (dot-dashed)
curves versus $\protect\rho $ in $\ z=8$ in the Kerr-AdS black hole
background with $l=5.$ Winding number is $N=1.$ }
\label{fig14}
\end{figure}

\end{document}